\documentclass[twocolumn,showpacs,preprintnumbers,amsmath,amssymb,nofootinbib]{revtex4-1}
\usepackage{graphicx}
\setcitestyle{round}

\usepackage{graphicx}
\usepackage{color}
\usepackage{bm}
\usepackage{braket}
\usepackage{amsmath}
\usepackage{natbib}
\usepackage{siunitx}
\usepackage{soul}
\usepackage{ulem}
\usepackage[version=4]{mhchem}

\newcommand{\ADD}[1]{\textcolor{black}{#1}}
\newcommand{\DEL}[1]{}

\begin{document}


\newcommand\blfootnote[1]{%
	\begingroup
	\renewcommand\thefootnote{}\footnote{#1}%
	\addtocounter{footnote}{-1}%
	\endgroup
}

\title{Growth of self-integrated atomic quantum wires and junctions of a Mott semiconductor}
\author{ Tomoya\,Asaba$^{1*\dagger}$, Lang\,Peng$^{1*}$, Takahiro\,Ono$^1$, Satoru\,Akutagawa$^{1}$,  Ibuki\,Tanaka$^1$,  Hinako\,Murayama$^{1,2}$, Shota\,Suetsugu$^{1}$, \ADD{Aleksandar\,Razpopov$^{3}$},  Yuichi\,Kasahara$^{1}$, Takahito\,Terashima$^{1}$, Yuhki\,Kohsaka$^1$, Takasada\,Shibauchi$^{4}$, Masatoshi\,Ichikawa$^1$, \ADD{Roser\,Valent\'i$^{3}$,} Shin-ichi\,Sasa$^1$, and  Yuji\,Matsuda$^{1\dagger}$}

\affiliation{
$^1$Department of Physics, Kyoto University, Kyoto 606-8502, Japan.\\
$^2$RIKEN Center for Emergent Matter Science, Wako, Saitama 351-0198, Japan\\
\ADD{$^3$Institut f\"ur Theoretische Physik, Goethe-Universit\"at, 60438 Frankfurt am Main, Germany}\\
$^4$Department of Advanced Materials Science, University of Tokyo, Kashiwa, Chiba 277-8561, Japan\\
$^{\dagger}$Correspondence to: asaba.tomoya.4t@kyoto-u.ac.jp and matsuda@scphys.kyoto-u.ac.jp\\
}

\date{\today}

\blfootnote{$^*$ denotes equal contribution.}
\blfootnote{$^{\dagger}$ denotes corresponding authors.}
\begin{abstract}
Continued advances in quantum technologies rely on producing nanometer-scale wires. Although several state-of-the-art nanolithographic technologies and bottom-up synthesis processes have been used to engineer such wires,  critical challenges remain in growing uniform atomic-scale crystalline wires and constructing their network structures.  Here we discover a simple method to fabricate atomic-scale wires with various arrangements, including stripes, X-, Y-junctions, and nanorings. Single-crystalline atomic-scale wires of a Mott insulator, whose band gap is comparable to those of wide-gap semiconductors, are spontaneously grown on graphite substrates by pulsed-laser deposition. These wires are one-unit-cell-thick and have an exact width of two- and four-unit-cells (1.4 and 2.8\,nm) and lengths up to a few $\mu m$.  
We show that the non-equilibrium reaction-diffusion processes may play an essential role in atomic pattern formation.  Our findings offer a new perspective on the non-equilibrium self-organization phenomena on an atomic scale, paving a unique way for the quantum architecture of nano-network.

\end{abstract}

\maketitle

\section*{Introduction}

The fundamentals of device technology dramatically change by the reduction of the dimensions. As the device size is reduced to the nanometer scale, the fabrication and integration of one-dimensional (1D) wire patterns become increasingly complicated and demanding. For the top-down approaches by using modern technologies with large-scale equipment, such as electron-beam and focused ion-beam lithography~\cite{liu2013top,lin2014flexible,dyck2019atom}, fabricating patterns of nanowires with thickness and width less than \SI{\sim 10}{nm} is technically challenging.  On the other hand, bottom-up technologies by using self-assembly processes~\cite{holmes2000control,mallet2008growth,kibsgaard2008atomic} suffer from controlling the uniformity of the wires.  Moreover, in the bottom-up methods,  the integration of nanowire arrays consists of two complicated steps: growing randomly-oriented nanowires and aligning them into an array. Most of the recent growth of nanowires is based on the vapor-liquid-solid method~\cite{wagner1964vapor,johansson2011recent,garnett2019introduction}. The assembly of the nanowires harnesses various processes such as Langmuir-Blodgett assembly~\cite{whang2003large,liu2010mesostructured}.  Therefore, the situation may call for a novel technology based on a fundamentally distinct concept that can fabricate uniform atomic-scale wires and engineer their nano-patterns.  

Apart from the semiconductor technologies, atomic quantum wires of electron systems open a new route to realizing exotic phases of matter. For example, it has been proposed that in the proximity of a supercondcutor, atomic quantum wires can host Majorana fermions, which can provide an approach to constructing a fault-tolerant quantum computer~\cite{kitaev2006anyons}.  Also, the quantum wires consisting of correlated electrons provide a platform for the physics of 1D non-Fermi liquid described by such as Tomonaga-Luttinger model, in which spin-charge separation plays an important role~\cite{giamarchi2003quantum}.  Furthermore, when the electron correlations are strong enough in the half filled band,  Mott insulators that can be seen as 1D quantum magnets are realized.  An exotic example of such a system is an Ising magnet forming a tri-junction (Y-junction), which has an exactly solvable ground state hosting Majorana fermions~\cite{tsvelik2013majorana}. Therefore, the fabrication of quantum wire arrays and junctions in correlated systems has the potential to significantly impact the field of condensed matter physics.

\begin{figure*}[t]
	\centering
	\includegraphics[width=\textwidth]{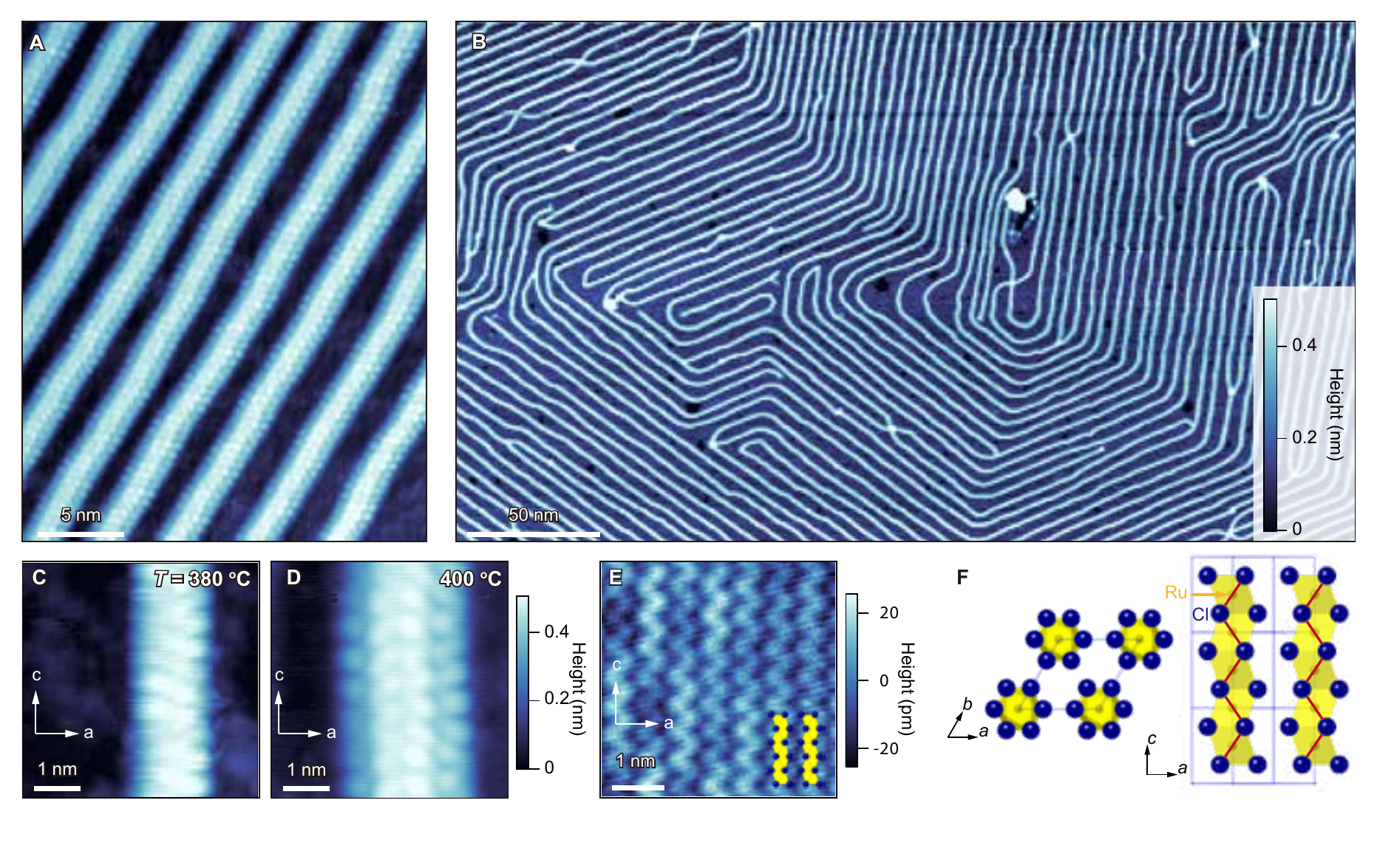}
	\caption{\label{Fig:basic}
        \textbf{Topographic images of $\beta$-RuCl$_3$ atomic-scale wires grown on HOPG surfaces.}
        \textbf{A,}  Topographic images highlighting atomic structures of the $\beta$-RuCl$_3$ wires consisting of four- $\beta$-RuCl$_3$ single-crystalline chains.  Periodic white spots represent chlorine atoms.  \ADD{The deposition temperature is \SI{400}{\degreeCelsius}.} The color scale is shared by \textbf{A} and \textbf{B}. The images are taken at \SI{2}{V} and \SI{30}{pA}.
        \textbf{B,} A topographic image of $\beta$-RuCl$_3$ on HOPG taken at \SI{3}{V} and \SI{20}{pA}.  Bright lines represent single-crystalline $\beta$-RuCl$_3$ wires with four-unit-cell-width and dark blue areas represent $a$-Ru-Cl, an amorphous material consisting of Ru and Cl. The deposition temperature is \SI{400}{\degreeCelsius}. 
        \ADD{
        \textbf{C, D,} Topographic images highlighting atomic structures of the $\beta$-RuCl$_3$ wires consisting of two-(\textbf{C}) and four-(\textbf{D}) $\beta$-RuCl$_3$ single-crystalline chains.  The deposition temperatures are \SI{380}{\degreeCelsius} (\textbf{C}) and \SI{400}{\degreeCelsius} (\textbf{D}). The color scale is shared by \textbf{C} and \textbf{D}. The images are taken at \SI{2}{V} and \SI{50}{pA} for \textbf{C}, and \SI{2}{V} and \SI{30}{pA} for \textbf{D}.
        }
        \textbf{E,} A topographic image of 2D monolayer $\beta$-RuCl$_3$ taken at \SI{3}{V} and \SI{50}{pA}.  Zigzag chains of chlorine atoms are arranged in parallel.
        \textbf{F,} Crystal structure of $\beta$-RuCl$_3$ viewing from directions normal to $ab$- (left) and $ac$- (right) planes.  The blue dashed lines denote the unit cell. In the right panel, the monolayer crystal structure is shown.  Zigzag red lines correspond to the zigzag chains of Cl atoms in \textbf{E}.
        } 
\end{figure*}

Here we demonstrate that uniform and long single-crystalline wires of \ce{\beta-RuCl3} in an atomic scale are reliably fabricated by a simple deposition technique. Furthermore, we manufactured several characteristic patterns pivotal for realizing quantum nano-circuits, including atomically-smooth junctions and nanorings. \ce{\beta-RuCl3} is an interesting material because it is a Mott insulator in which electron-electron interaction opens a gap, while at the same time its band gap is comparable to those of wide-gap semiconductors. The formation of these nanowire pattern takes place as a part of the thin film growth, but the growth process is essentially different from the conventional ones.  In the present process, the formation and integration of nanowires are achieved simultaneously, making this method unique and attractive for applications.  Furthermore, we obtain multiple pieces of evidence that the observed atomic-scale wire patterns emerge as a result of a characteristic self-organization process.  We will discuss that the non-equilibrium reaction-diffusion processes may play an important role for the uniformly aligned patterns. We will further point out a possible atomic-scale Turing mechanism~\cite{TuringOriginal}, which has been  discussed for macroscopic pattern formations in chemistry~\cite{horvath2009experimental}  and biology~\cite{kondo1995reaction,sick2006wnt,sheth2012hox}.

\section*{Results}

\begin{figure*}[t]
	\centering
	\includegraphics{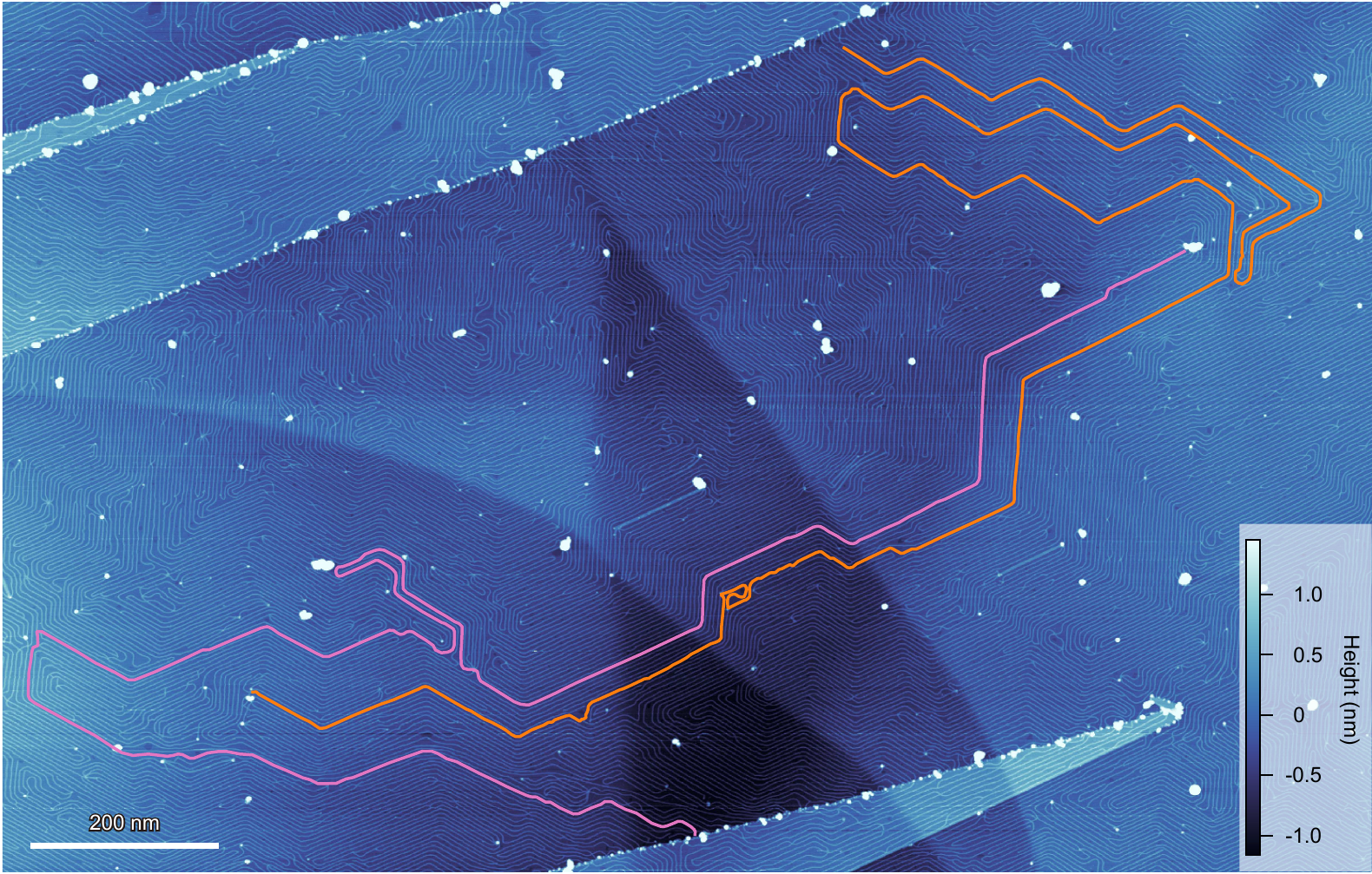}
	\caption{\label{Fig:long}
        \textbf{Topographic image of $\beta$-RuCl$_3$ atomic wires extending over a few micrometers.}
        The orange and magenta lines are overlaid on atomic wires of $\beta$-RuCl$_3$ with a four-unit-cell width (\SI{\sim 2.8}{nm}).  Their lengths are longer than \SI{3}{\micro\meter}.  \ADD{The high clusters are objects adhering to the surface probably during the growth process.}  The topographic image was taken at \SI{3}{V} and \SI{20}{pA}.
        } 
\end{figure*}

 We evaporated high-quality \ce{RuCl3} on the surfaces of a highly-oriented-pyrolytic-graphite (HOPG) by the pulse-laser-deposition (PLD) method  (see Methods).  Figures\,\ref{Fig:basic}A-E show the scanning tunneling microscope (STM) topographic images of the growth results.  Shown in Fig.\,\ref{Fig:basic}A is an atomic\ADD{-}resolution \ADD{image of a sample} grown at the deposition temperatures of \SI{400}{\degreeCelsius}.  We find that the surface is covered by a unique pattern of wires (bright lines), which are ordered and almost evenly spaced.  Each wire consists of periodically spaced atoms, demonstrating the single-crystalline structure.  Figure\,\ref{Fig:basic}B  displays a typical large-scale picture of the growth\ADD{, revealing that the wires grow in some preferred directions with being connected smoothly at the corners.}  \ADD{The direction is parallel to the zigzag-direction of the carbon honeycomb lattice of the substrate (orange arrow in Fig.\,\ref{Fig:stripe}B).  Reflecting the crystal structure of HOPG, the growth direction of wires is oriented \ang{60} away from each other.}  The regions between the atomic wires (\ADD{dark }blue areas in Figs.\,\ref{Fig:basic}A and \ref{Fig:basic}B) are filled with a thin material.  Since no periodic lattice structure is seen \ADD{(Fig.\,\ref{Fig:stripe}A)}, the thin material is in an amorphous form consisting of Ru and Cl, denoting as $a$-Ru-Cl.

\begin{figure*}[t]
	\centering
	\includegraphics{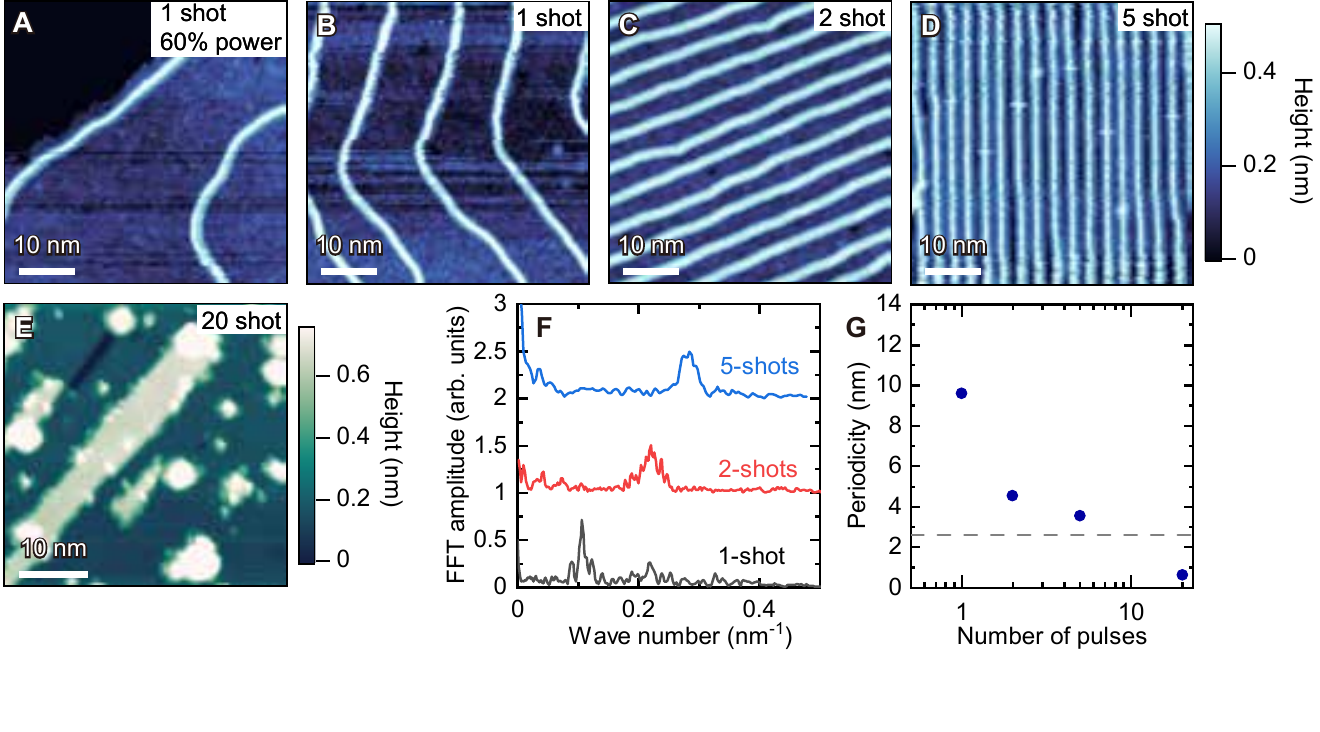}
	\caption{\label{Fig:tuning}
        \textbf{Stripe patterns of $\beta$-RuCl$_3$ atomic-scale wires.}
        \textbf{A-D,} Topographic images of $\beta$-RuCl$_3$ wires with four-unit-cell width grown at \SI{400}{\degreeCelsius}. By changing the deposition time of the laser from 1 to 5 shots, the wire distance can be tuned  from much longer than \SI{10}{nm} (\textbf{A}) to shorter than \SI{2}{nm} (\textbf{D}).  The power of the laser pulse is further attenuated to 60 \% for \textbf{A}.  The color scale is shared by \textbf{A--D}. 
        \textbf{E,} A topographic image of a $\beta$-RuCl$_3$ monolayer thin film grown by a further increase of the deposition time to 20 shots. Green and white regions correspond to mono- and double-layer thick $\beta$-RuCl$_3$, respectively.  No 1D wire pattern is observed. The setpoint conditions are \SI{20}{pA} and \SI{3}{V} (\textbf{A, B, E}), \SI{30}{pA} and \SI{3}{V} (\textbf{C, D}).
        \ADD{
            \textbf{F,} Line profiles of FFT images in the direction of peaks corresponding to the wire repetition.  The curves are vertically shifted for clarity.
            \textbf{G,} The periodicity (the inverse of the wavenumber) is plotted as a function of the number of pulses.  The dashed gray line indicates the width of the 4-chain wire.  The data point for the 20-shot represents the lateral lattice constant of monolayer $\beta$-RuCl$_3$.
        }
        }
\end{figure*}

\begin{figure*}[t]
	\centering
	\includegraphics{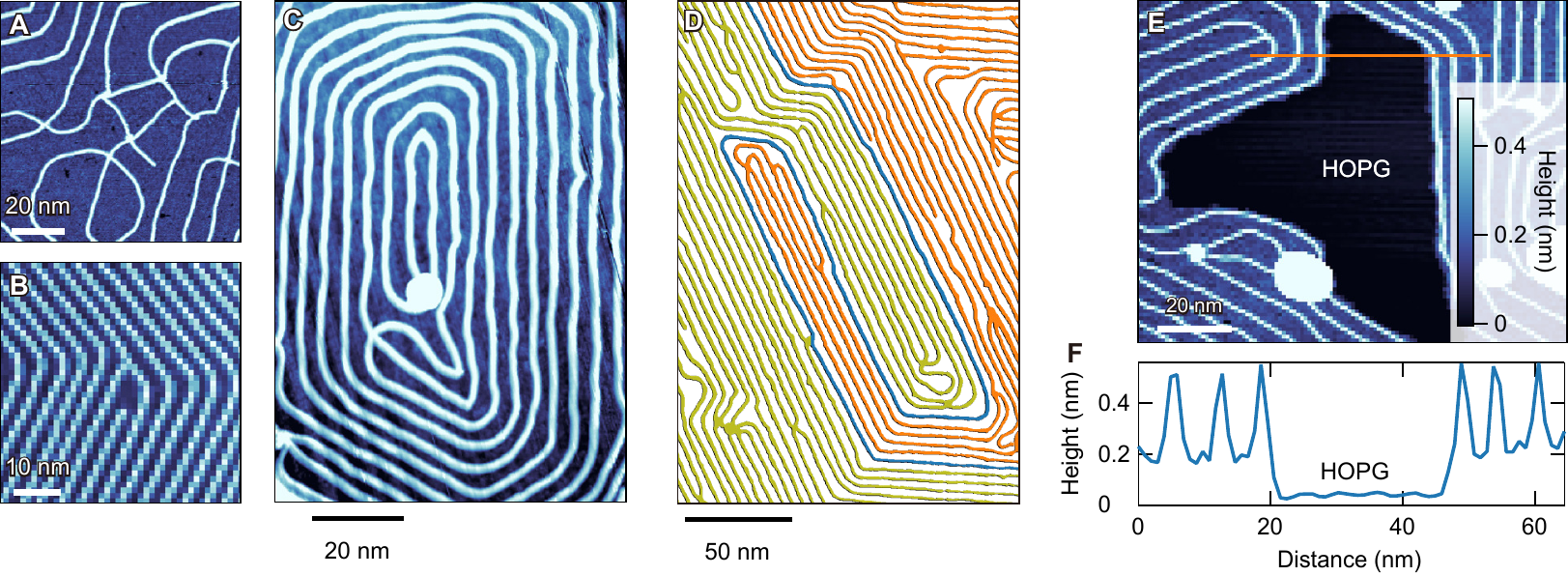}
	\caption{\label{Fig:evidence}
        \textbf{Topographic images exemplifying spatial patterns formed by reaction-diffusion processes.}
        \textbf{A,} Multiple crossings of atomic wires of $\beta$-RuCl$_3$ with four-unit-cell width observed in a region with very low wire density. The color scale is shared by \textbf{A-C} and \textbf{E}.
        \textbf{B,} Termination of a wire with four-unit-cell-width accompanied by a Y-shaped junction.
        \textbf{C,} Winding spiral pattern.
        \textbf{D,} Collision of growing waves of atomic wire colored in yellow and orange,  forming a spiral pattern. The boundary of the waves is colored in blue. The raw image is shown in  Fig.\,\ref{Fig:rawspiral}.
        \textbf{E,} A depleted region of $a$-Ru-Cl, where the HOPG surface appears.  Atomic-scale wire propagation terminates at the boundary of this region.
        \textbf{F,} Height profile along the orange line indicated in \textbf{E}. The setpoint conditions are \SI{20}{pA} and \SI{3}{V} (\textbf{A}), \SI{30}{pA} and \SI{2}{V} (\textbf{B}), \SI{30}{pA} and \SI{2}{V} (\textbf{C}), \SI{30}{pA} and \SI{3}{V} (\textbf{D}), and \SI{20}{pA} and \SI{3}{V} (\textbf{F}).
        } 
\end{figure*}

To identify the material forming the atomic-scale wires, we extend the deposition time to grow 2D monolayer and thicker films.  As shown by the STM image in Fig.\,\ref{Fig:basic}E,  we can obtain single-crystalline 2D monolayer films. No wires are formed on the monolayer film, enabling continuous layer-by-layer growth~\cite{frank1949one}. Based on the \ADD{ combination of the characteristic zigzag chains and} X-ray diffraction pattern of the thicker films (Fig.\,\ref{Fig:Xray}),  we identify the crystal \ADD{as} \ce{\beta-RuCl3}. The zigzag chains \ADD{originate from the topmost Cl} atoms \ADD{as highlighted} by the red lines in Fig.\,\ref{Fig:basic}F. \ADD{The periodicity along the chain corresponds to the lattice constant of \ce{\beta-RuCl3}, further corroborating that the chain is \ce{\beta-RuCl3}.} The atomic wires shown in Fig.\,\ref{Fig:basic}A have the same periodicity as the \ce{\beta-RuCl3} chain along the wire direction, and their widths coincide with double \ADD{(Fig.\,\ref{Fig:basic}C)} and quadruple \ADD{(Fig.\,\ref{Fig:basic}D)} \ce{\beta-RuCl3} chains. The same zigzag structure is resolved only in the inner chains of four-unit-cell-width wires.  \ADD{The apparent difference between the inner and outer chains may be caused by scanning an edge (the outer chain in this case) using a tip with finite curvature or may reflect the difference in the electronic states of the inner and outer chains.}  As discussed later ( Fig.\,\ref{Fig:evidence}F), the height of the wire is \SI{\sim 0.5}{nm}, which is consistent with the monolayer thickness of \ce{\beta-RuCl3}. 
The height of $a$-Ru-Cl is typically less than \SI{0.2}{nm}, much smaller than \ce{\beta-RuCl3}. 

\ADD{A salient feature of the wire is its length.}  As demonstrated in Fig.\,\ref{Fig:long}, the length is as long as more than \SI{3}{\micro\meter} (a length-width ratio is more than 1000), and both the width and height of the wire are quite uniform over its entire length.  This length is significantly longer than most of the nanowires with a few atomic widths reported so far, which are typically \numrange[range-phrase = --]{10}{100} nanometer lengths.  To our knowledge, the presently observed patterns of uniform 1D atomic wires with mesoscipic length scale are unique and unprecedented.  

Based on the above results, we conclude that the atomic wires consist of two or four \ce{\beta-RuCl3} single-crystalline chains \ADD{growing in the [001]-direction} directly on the HOPG surfaces.  Importantly, at the deposition temperature of \SI{400}{\degreeCelsius}, the wires always consist of quadruple chains of \ce{\beta-RuCl3} \ADD{ (Fig.\,{\ref{Fig:basic}D}) while the width is reduced to double chains at the lower deposition temperature of \SI{380}{\degreeCelsius} (Fig.\,{\ref{Fig:basic}C})}.  Hereafter, we denote the mechanism by which four (two) chains bundle together to form a single wire ``four- (two-) chain rule.''  The detailed mechanism of the four (two)-chain rule is not apparent, and exploring its quantum-mechanical origin deserves future studies. However, this rule is crucial for building a model, as discussed later.    We note that the $c$-axis lattice constant of \ce{\beta-RuCl3} does not match the lattice constants and their integer multiple of the carbon honeycomb lattice. Therefore, the epitaxial coupling with HOPG is not essential for the 1D wire formation, further supported by the curved 1D wires shown in Figs.\,\ref{Fig:tuning}A  and \ref{Fig:evidence}A. 

\begin{figure*}[t]
	\centering
	\includegraphics{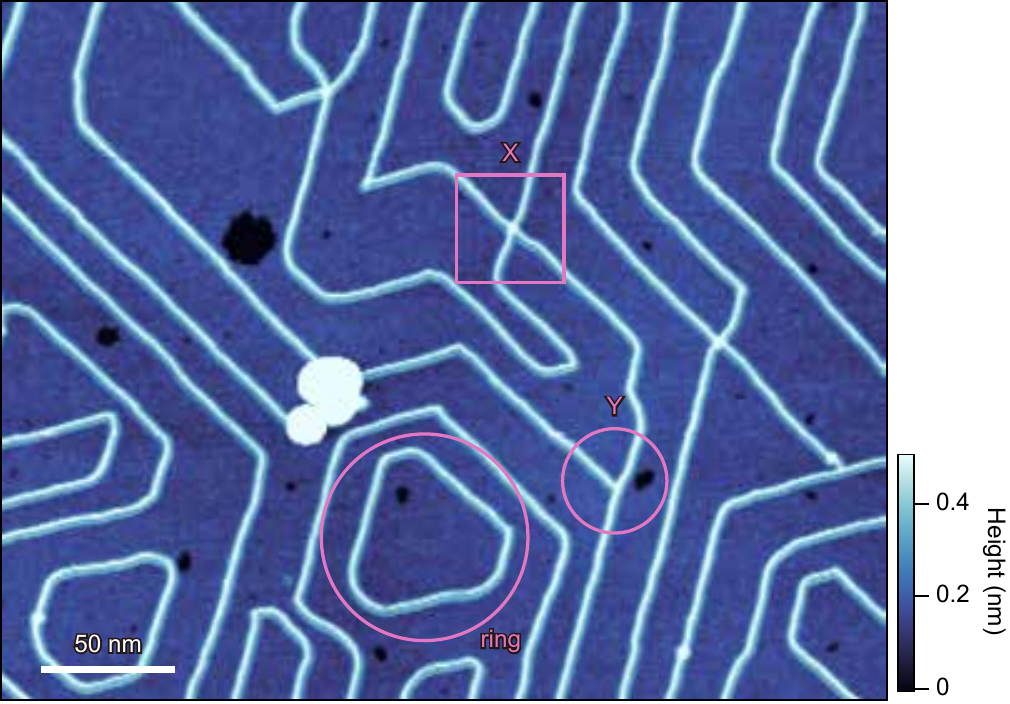}
	\caption{\label{Fig:pattern}
        \ADD{\textbf{Spontaneous formation of junctions and rings.}
        A topographic image taken at \SI{20}{pA} and \SI{3}{V} shows X-, Y-junctions, and rings of four-unit-cell wide $\beta$-RuCl$_3$ wire. }
        }
\end{figure*}

\begin{figure*}[t]
	\centering
	\includegraphics{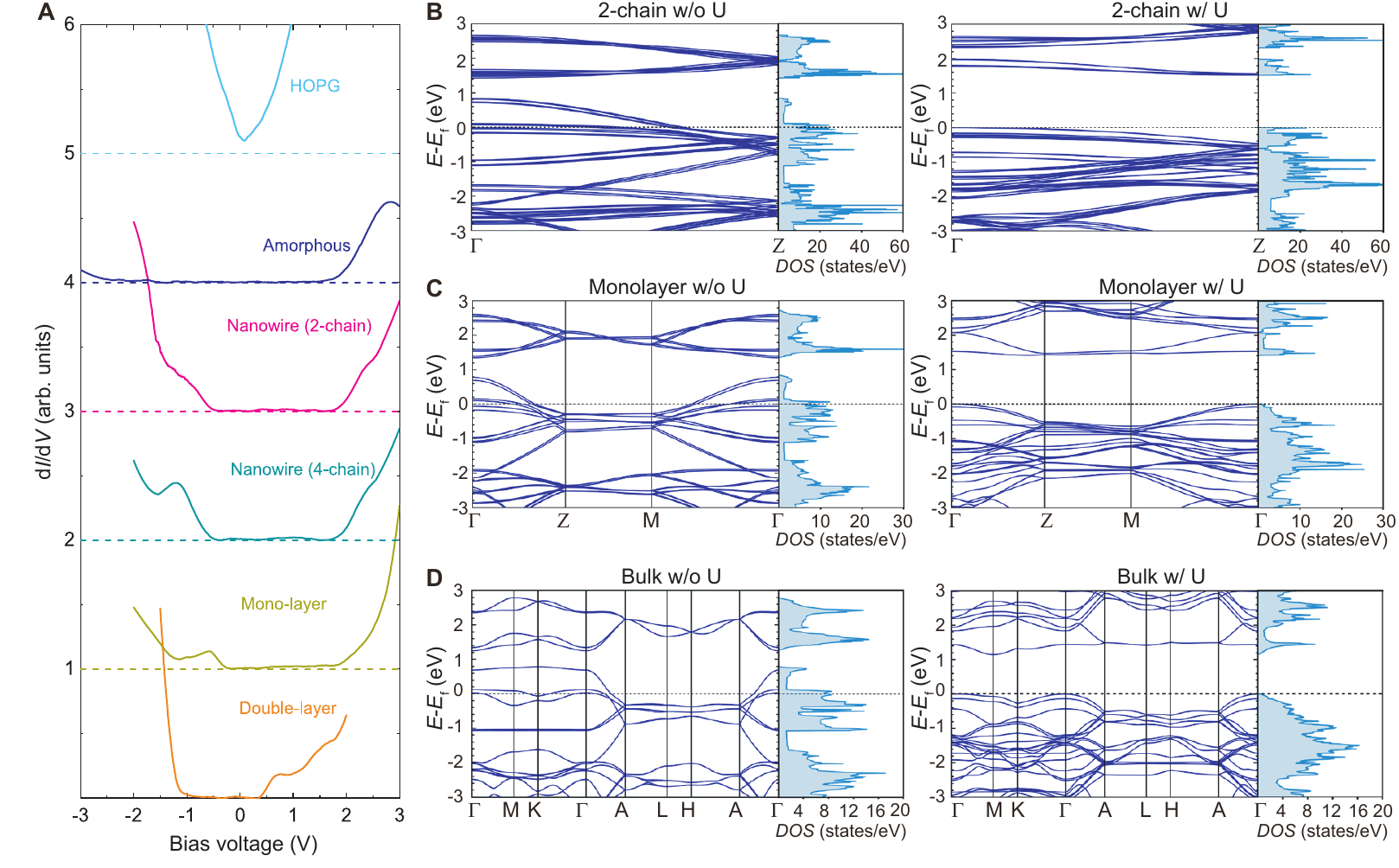}
	\caption{\label{Fig:dIdV}
       \ADD{
           \textbf{Electronic structures of $\beta$-RuCl$_3$ atomic-scale wires.}
           \textbf{A,} d$I$/d$V$ spectra of $\beta$-RuCl$_3$.  The setpoint conditions are \SI{2}{V} and \SI{100}{pA} (double-layer), \SI{3}{V} and \SI{100}{pA} (mono-layer and 2-chain nanowire), \SI{3}{V} and \SI{50}{pA} (4-chain nanowire and amorphous), and \SI{1}{V} and \SI{100}{pA} (HOPG).  The spectra of 4-chain nanowire and amorphous are doubled so that the setpoint current is equivalent to the others.  The curves are vertically shifted for clarity, and each dashed line represents the base line.  We did not specify a chain in a wire due to the piezo creep being relatively large at \SI{78}{K}.
           \textbf{B-D,} The calculated electronic band structure of 2-chain wire, monolayer, and bulk $\beta$-RuCl$_3$, respectively. The left (right) columns show the calculations without (with) electronic correlations.}
        }
\end{figure*}

As shown in Figs.\,\ref{Fig:tuning}A--D, the spacing between the 1D wires is widely tunable ranging from lengths  longer than \SI{10}{nm} (Fig.\,\ref{Fig:tuning}A) to less than \SI{2}{nm} (Fig.\,\ref{Fig:tuning}D), by changing the deposition time. The longer deposition time results in the formation of the 2D monolayer \ce{\beta-RuCl3} thin film (Fig.\,\ref{Fig:tuning}E).  \ADD{A more quantitative analysis using the fast Fourier transform is shown in Fig.\,\ref{Fig:tuning}F.  The periodicity of the wire, the inverse of the wavenumber at the peak position in Fig.\,\ref{Fig:tuning}F, monotonically decreases as increasing the deposition time  (Fig.\,\ref{Fig:tuning}G).  For the 5-shot, the periodicity is \SI{3.5}{nm}, which is merely above the width of wires (\SI{\sim 2.8}{nm}), indicating that the wire separation is less than \SI{1}{nm}.}

\ADD{Another few intriguing patterns found} in addition to the stripe patterns \ADD{are} X-, Y-junctions as well as nano-rings composed of atomic-scale wires, as indicated in Fig.\,\ref{Fig:pattern}.    The height of the nano-ring and Y-junction is uniformly one-unit-cell.  These results demonstrate the fabrication of atomically smooth junctions and rings without introducing defects and clusters, which are crucially important for nanocircuits.

\ADD{To reveal the electronic structures, we measured} the differential tunneling conductance (d$I$/d$V$) spectra (Fig.\,\ref{Fig:dIdV}A), which are proportional to the local density of states, on wires with two- and four-unit-cell width.  For the comparison, spectra of 2D \ce{\beta-RuCl3} with mono and double layer thicknesses (Fig.\,\ref{Fig:tuning}E), $a$-Ru-Cl, and HOPG are also shown.  In contrast to semimetallic HOPG with a linear dispersion, clear gap structures are seen in all \ce{\beta-RuCl3}, indicating the semiconducting or insulating electronic structures.  
\ADD{To unveil the origin of the energy gap, we performed systematic band calculations of a 2-chain wire, a monolayer, and bulk \ce{\beta-RuCl3}, considering electron correlations and spin-orbit interactions (Fig.\,\ref{Fig:dIdV}B-D).  All the cases are metal with a finite density of states at the Fermi energy unless the electron correlation is included, whereas it is turned on, an energy gap opens at the Fermi energy for all cases, indicating that \ce{\beta-RuCl3} is a Mott insulator.}  

\begin{figure}[t]
	\centering
	\includegraphics[width=\linewidth]{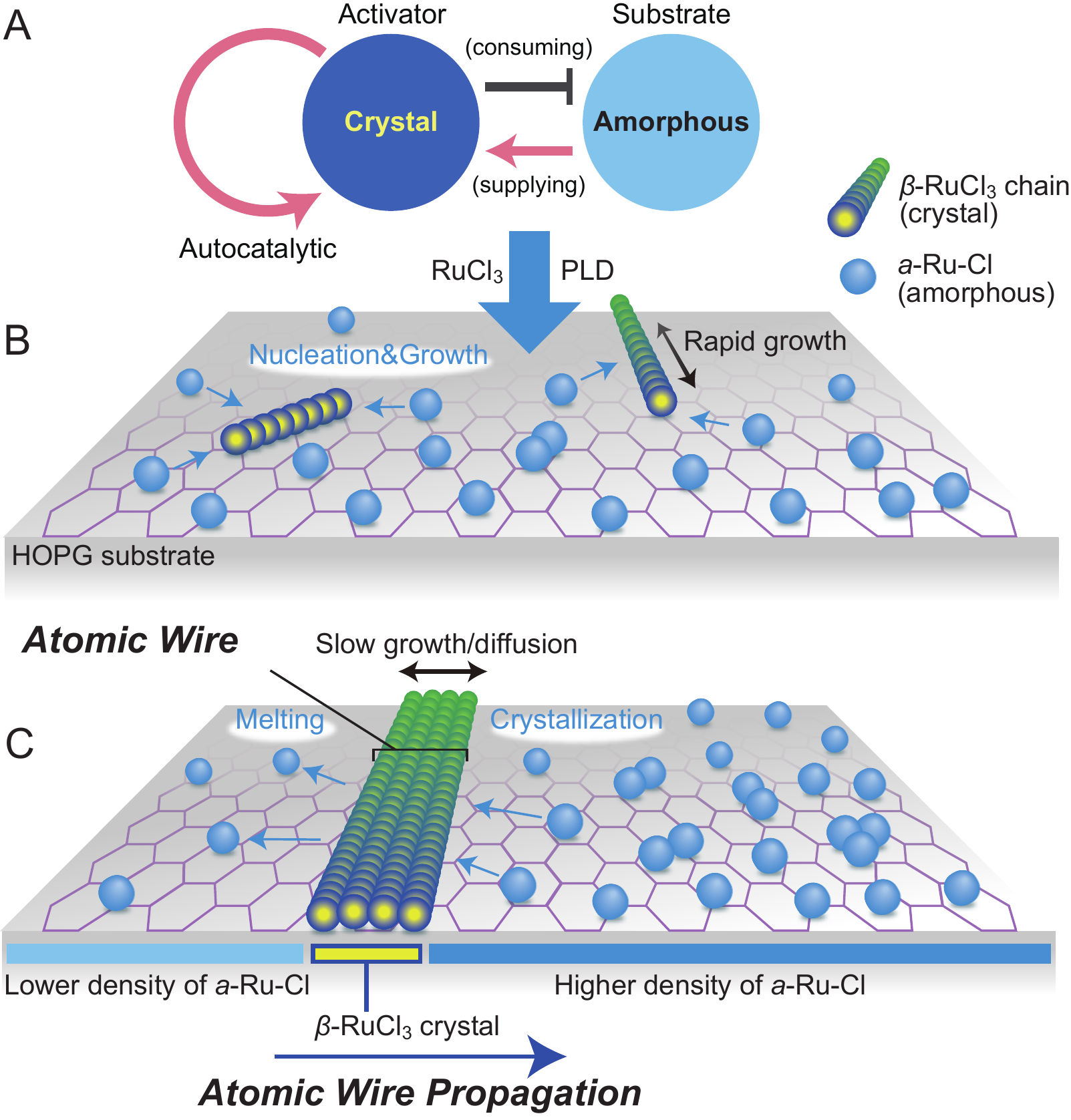}
	\caption{\label{Fig:concept}
        \textbf{Schematic diagrams of the atomic-wire formation by Turing mechanism.}
        \textbf{A,} Activator-depleted substrate scheme. Depletion of the substrate acts as an inhibitor in the conventional activator-inhibitor system in the Turing mechanism.
        \textbf{B, C,} Crystal growth and diffusion process of $\beta$-RuCl$_3$.  The chemical reaction process occurs on both sides of the 1D wires to form and decompose $\beta$-RuCl$_3$, but the reaction is activated more frequently on the side with a higher concentration of $a$-Ru-Cl. The atomic wires propagate towards the direction with a higher concentration of $a$-Ru-Cl. This process describes the reaction-diffusion origin of the pattern formation.
        } 
\end{figure}
 
Considering Figs.\,\ref{Fig:tuning}A-E, the formation of the nanowire array appears to be part of the thin-film growth. However, it should be emphasized that this crystal growth process is essentially different from any of the growth processes previously known: island, layer-by-layer, and their combination~\cite{volmer1926keimbildung,frank1949one,stranski1937theorie}.  It is interesting to note that the spontaneous formation of distinctive stripe patterns has been reported in macroscopic and mesoscopic scales in various chemical and biological systems. Two contrasting approaches have been advocated to explain the stripe pattern formation in the previous systems.  One is an equilibrium process,  in which the periodic spatial organization appears as a result of the static potentials consisting of competition between attractive and repulsive interactions with two different length scales or purely repulsive isotropic interactions~\cite{seul1995domain,malescio2003stripe,de2006columnar}.  The other is to view the 1D stripe pattern as the result of a dissipative structure that exhibits complex non-equilibrium reaction-diffusion.

To elucidate the mechanism of the present pattern formation, noteworthy about the observed patterns is the emergence of several distinct characteristic features displayed in Figs.\,\ref{Fig:evidence}A-E, in addition to the stripe patterns.  It should be stressed that  from these peculiar patterns, static interactions are highly unlikely to be the driving force of the atomic wire array formation.  As shown in Fig.\,\ref{Fig:evidence}A,  when the distance between the wires becomes longer,  the stripe patterns disappear, but the wire crossings still remain. These results contradict the pattern formation driven by static interactions, whether repulsive or attractive, because crossings typically induce energy loss if static interactions drive the pattern formation. If attraction-driven, it is more stable to condense into a 2D phase separation structure, i.e., crystalline domains. In the present case, such a crystallized domain has not been observed. For the repulsion-driven case, the uniformly dispersed pattern is more stable, which contradicts the present case with dense and sparse regimes coexisting, as shown in Fig.\,\ref{Fig:evidence}A.  Thus, the possibility of the static interactions can be ruled out as the origin of the atomic wire patterns.

We point out that non-equilibrium reaction-diffusion processes may play an essential role in forming the observed peculiar atomic-wire patterns.  Indeed, in reaction-diffusion systems, the crossings are allowed to form~\cite{cross1993pattern}.  In addition, as depicted in Fig.\,\ref{Fig:evidence}B, when the number of wires changes, the Y-junction  accompanied by an edge dislocation appears.  For a stripe pattern, the number of lanes' discordance is typically resolved in two ways. One is to narrow the distance between the lanes. The other one is to form transiently or stationary emerging Y-junctions, edges of lanes, and shrinking lanes, which we observed in Fig.\,\ref{Fig:evidence}B. Such a pattern is a typical fingerprint for the reaction-diffusion process~\cite{TuringOriginal,nicolis1977self}.   
	
Moreover, the presence of the phase singularity, an unique feature in reaction-diffusion systems~\cite{kuramoto1984chemical}, is demonstrated by the spiral patterns in Figs.\,\ref{Fig:evidence}C and \ref{Fig:evidence}D. Figure \ref{Fig:evidence}C shows a typical winding spiral pattern (see also  Fig.\,\ref{Fig:NewSpiral}).  
The pattern in Fig.\,\ref{Fig:evidence}D is created as a result of two waves, one colored yellow and the other orange, propagating from different directions and colliding with each other. We emphasize that these spiral patterns cannot be generated in the equilibrium systems, providing strong evidence for the non-equilibrium process~\cite{cross1993pattern, kuramoto1984chemical} (see also SI). Furthermore,  when the deposition temperature is lowered to \SI{380}{\degreeCelsius},  the HOPG surface with no $a$-Ru-Cl appears as shown by the black area in Fig.\,\ref{Fig:evidence}E. The wires do not propagate through a depletion region by themselves without accompanying $a$-Ru-Cl regions nearby, which is confirmed by the height profile (Fig.\,\ref{Fig:evidence}F) along the orange line in Fig.\,\ref{Fig:evidence}E. The absence of the propagation of wires to the depletion region is consistent with the reaction-diffusion picture because it suggests that the production of atomic-scale wires regulates the concentration of the $a$-Ru-Cl and vice versa.

We note that the STM is too slow to capture the dynamic process of the thin-film growth. Therefore, even though these STM results support the reaction-diffusion process as the origin of the atomic wire pattern formation, measurements that can directly detect dynamical processes in atomic-scale are strongly desired to fully understand the growth mechanism.

\section*{Discussion}

Assuming that the reaction-diffusion process is the origin of the pattern formation of atomic wires, it is tempting to speculate the stripe pattern to be a manifestation of Turing instability~\cite{TuringOriginal}.  This instability is  the most prominent mechanism in many different classes of self-organization processes, leading to the spontaneous emergence of spatially periodic patterns~\cite{TuringOriginal}.  In the following, we discuss the possibility that the observed patterns are associated with the Turing instability. 

Turing patterns occur when the diffusion coefficient of an inhibitor is significantly larger than that of an activator; local activation and lateral inhibition destabilize uniform stationary states. After the first experimental realization of Turing patterns around 1990~\cite{castets1990experimental,ouyang1991transition},  40 years after Turing's prediction, the Turing mechanism has been recognized as a significant driving force that can induce the chemical~\cite{horvath2009experimental}  and biological~\cite{kondo1995reaction,sick2006wnt,sheth2012hox} pattern formations.  However, despite its applicability to various systems with a wide length scale ranging from more than centimeters to submicrons~\cite{tan2018polyamide}, the atomic-scale formation of the Turing pattern down to a few nanometers has been seldom examined.  Very recently, striped wrinkle patterns of the 2D bismuth atoms on NbSe$_2$ have been discussed in the analogy of the Turing pattern by associating the atomic displacements with activator and inhibitor~\cite{fuseya2021nanoscale}. However,  further evidence is required to clarify whether these patterns appear due to the non-equilibrium self-organization process because similar patterns can be induced even by an equilibrium process stemming from the static interactions~\cite{seul1995domain,malescio2003stripe,de2006columnar}. In addition, the present finding of the atomic-scale wire fabrication is distinct from these wrinkle patterns on 2D continuous films, which is essentially a surface reconstruction phenomenon.  Thus,  the mechanism that underlies diffusion-driven instability on the atomic scale, where quantum mechanics is essentially important, can be fundamentally different from that on the larger scale. Moreover, it is an open question whether the large differences in diffusion rates of reacting species can be satisfied on the atomic scale.
  
Despite that the Turing mechanism in atomic scale is largely unexplored, we point out that  some of the observed patterns appear to be consistent with this instability.  Let us deduce an underlying model that describes the present system following the Turing instability. The emergence of the depletion areas shown in Fig.\,\ref{Fig:evidence}E suggests the activator-depleted substrate scheme~\cite{koch1994biological}, where the `substrate' denotes the chemically reacting substance.  As schematically shown in Fig.\,\ref{Fig:concept}A, \ce{\beta-RuCl3} crystal is an activator that enhances nucleation growth and consumes the substrate $a$-Ru-Cl, and the depletion of the substrate acts as an inhibitor in the conventional activator-inhibitor model for the Turing process.  We illustrate the crystal growth and diffusion process in Figs.\,\ref{Fig:concept}B and \ref{Fig:concept}C.  The crystal growth of 1D \ce{\beta-RuCl3} wires proceeds along the chain ([001]) direction because of the  quasi-1D crystal structure of \ce{\beta-RuCl3}.

Since the reaction between the wire and amorphous is activated more frequently on the side with a higher concentration of amorphous, the wires propagate towards the direction with a higher concentration of $a$-Ru-Cl, as illustrated in Fig.\,\ref{Fig:concept}C.  In a typical crystal growth, the crystal reacts with a-Ru-Cl, resulting in the lateral growth (layer-by-layer growth). However, in the present case, the atomic wires do not laterally grow but rather propagate towards the direction with a higher concentration of $a$-Ru-Cl due to the four-chain rule. The diffusion rate of the \ce{\beta-RuCl3} crystal is significantly slower than that of $a$-Ru-Cl because of the significant difference between particle hopping rates in them;  The diffusion rate is proportional to $\sqrt{N}$, where $N$ is the size of the substance. Since the typical volume of \ce{\beta-RuCl3} wire is more than 100 times larger than that of a-Ru-Cl, the diffusion rate of \ce{\beta-RuCl3} is expected to be more than ten times smaller than a-Ru-Cl. Note that the volume of a-Ru-Cl is $\sim$ 0.2 x 0.2 x 0.2 nm$^3$, while that of the wire is 2 x 0.6 x L nm$^3$, where $L$ is the length of the wire. $L$ is typically \ADD{longer} than \SI{100}{nm} in the present case. This suffices the necessary condition for the Turing instability that the diffusion coefficient of the inhibitor is significantly larger than that of the activator. As shown in Fig.\,\ref{Fig:Simulation}, a reaction-diffusion model describing the activator-substrate system exhibits a stripe pattern through the Turing instability~\cite{TuringOriginal} (see SI).  Also, we note that while the uniform stripe pattern appears when the density of atomic wires is high (as shown in Figs.\,\ref{Fig:tuning}C and \ref{Fig:tuning}D), the pattern is not periodic when the density of atomic wires is low (Figs.\,\ref{Fig:pattern} and \ref{Fig:evidence}A). We point out that this non-periodic pattern does not rule out the possibility of Turing mechanism as the origin of the uniform pattern at high-density (see SI). 
 
The present method of growing uniform atomic wires also has the potential to be applied to a wide variety of systems and opens up new avenues for the future of nanowire technology.  The robust and flexible growth of wires on the atomic terraces has great advantages for the applications.  Applications include the use of atomic patterns as lithographic masks and the use of the nanowires themselves.  We also point out that the present results are important for the basic science.  \ADD{As shown in Fig.\,\ref{Fig:dIdV}, \ce{\beta-RuCl3} is a Mott insulator with $J_{\rm eff}=1/2$,} in which the electron correlation \ADD{is predominant for the low-energy electronic states.}

We note that the gap formation due to Peierls instability is excluded because no dimerization is observed in the STM images.  In this case, the present \ce{\beta-RuCl3} wires provide a platform for the physics of pure 1D quantum magnets described by such as Tomonaga-Luttinger liquid and Haldane gap systems~\cite{lacroix2011introduction}.  Also, it has been known that another polymorphic form \ce{\alpha-RuCl3} with 2D honeycomb lattice is a candidate material of a Kitaev quantum spin liquid hosting Majorana fermions~\cite{kasahara2018majorana,yokoi2021half,biswas2019electronic}, which can provide an approach to constructing a fault-tolerant quantum computer~\cite{kitaev2006anyons}.  Combined with \ce{\alpha-RuCl3}, the atomic-scale wires and junctions of \ce{\beta-RuCl3} might be useful for quantum circuits that utilize Majorana fermions.  Furthermore, nanowire fabrication could be realized by the same mechanism in materials with 1D structure similar to \ce{\beta-RuCl3}. \ADD{For instance, transition metal trihalides with the same or similar quasi-1D crystal structure as \ce{\beta-RuCl3} have recently been focused on, and our method may be applied to this family~\cite{fu2022transition}. Other exciting candidates with quasi-1D crystal structures include transition metal chalcogenides such as WTe$_2$~\cite{ali2014large} and $M_6X_6$~\cite{shang2020atomic}.}

The fabrication of nanowires and junctions has the potential to significantly increase the integration of electronic circuits.  The present 1D crystalline atomic wires and their networks provide  fascinating physical playgrounds of non-equilibrium self-organization phenomena in atomic scale, exotic electronic states and  quantum technologies. 

\section*{Materials and methods}
\textbf{Sample growth:}
Atomic-scale wires and thin films of \ce{\beta-RuCl3} were grown on the surfaces of HOPG by the PLD method.  To obtain a fresh surface, the HOPG was cleaved, followed by annealing at \SI{400}{\degreeCelsius} for \SI{15}{min}.  The pressure of the PLD chamber was kept at about \SI{d-5}{Pa} during the deposition.  An \SI{81}{mJ} pulsed yttrium-aluminum-garnet laser operating at a wavelength of \SI{1064}{nm} was used to vaporize a solid target of high purity \ce{\alpha-RuCl3} single crystals, which were synthesized by the chemical vapor transport method.  To grow the films with atomic layer thickness, a 16-18 ns pulsed laser was used, and the laser power was tuned by using the attenuator. We find that the $\beta$ phase of \ce{RuCl3} can be grown at HOPG temperature between 350 and 400\,$^{\circ}$C in a high vacuum.  The growth of \ce{\beta-RuCl3} was confirmed by X-ray diffraction analysis of the thicker films grown on the same deposition conditions.  The growth condition was controlled by changing the number of pulses, laser power, and HOPG temperature.  After the growth, the films were transferred by a portable vacuum chamber to the STM chamber to image topographies and measure the local electronic properties.\\

\noindent
\textbf{STM measurements:}
The STM experiments were performed with a custom-made Unisoku ultrahigh vacuum STM at \SI{78}{K}. The STM tips were electrochemically-etched tungsten wires cleaned by electron-beam heating and conditioned on Ag(111) surfaces evaporated on Si(111) surfaces. The bias voltage was applied to the sample.  All topographic images were taken in the constant-current mode.  The differential conductance was measured with the standard lock-in technique at \SI{973}{Hz} with a modulation voltage of \SI{20}{mV}.\\

\noindent
\ADD{
\textbf{DFT calculations:}
All calculations are performed on the experimentally obtained crystal structure taken from Ref~\cite{Fletcher_1967}.
The structure is in the space group 185  $P6_3cm$  with lattice parameters $a = b = 6.120$~\AA~and $c = 5.658$~\AA, where the Ru-Ru chains propagate in the c direction.
The Ru-Ru distance within the chain is 2.829~\AA~and there is no dimerization.
This is in agreement with the presented STM images in this work which do not observe any dimerization either.}

\ADD{The experimental structure has been additionally compared with the
fully relaxed structure obtained within DFT via the VASP simulation package~\cite{Kresse_VASP}~version 6.3.0.
As exchange-correlation functional
we considered the Generalized Gradient Approximation (GGA)~\cite{GGA} where we included the Coulomb correction via the Dudarev~\cite{Dudarev_LDA+U}  GGA+U scheme with an effective Coulomb repulsion $U_{\text{eff}} = 3.7$~eV.
The simulations have been performed with a planewave cut-off of 600 eV for the expansion of the basis set, and on a 8$\times$8$\times$6 k-mesh.
The relaxation is performed including spin-orbit effects and van der Waals corrections via the DFT+D2 method of Grimme~\cite{grimme_vdw}, using a ferromagnetic state on the Ru sites, until the forces for each atom in all directions decrease down to $10^{-3}$~eV/\AA.
The comparison shows that the relative distances between the atoms deviate by a small amount, approximately 0.02~\AA, with respect to the experimentally reported distances
and therefore we continue with the experimentally reported structure.}

\ADD{To calculate the electronic properties of the experimental system we use the full potential local orbital (FPLO)~\cite{fplo} package version 21.00-61 and the Generalized Gradient Approximation (GGA)~\cite{GGA} where we include the Coulomb correction for the strongly localized Ru 4d electrons via the GGA+U approximation using the atomic limit method~\cite{Koepernik_2009_atomic_limit}. We apply a Coulomb correction of $U = 4.1$~eV~and $J = 0.37$~eV, which corresponds to approximately $U_{\text{eff}} = 3.7$~eV.
This value is slightly higher than reported cRPA values for a related Ru-system, the $\alpha$-RuCl$_3$ compound~\cite{kaib2022electronic}. Since $\beta$-RuCl$_3$ consists of quasi 1D chains we expect that the screening effects will be slightly reduced and thus the effective Coulomb repulsion should be slightly larger.  All calculations are spin polarized with ferromagnetically aligned magnetic moments on the Ru sites. Due to the heavy Ru atoms we perform fully relativistic calculations with a spin quantization axis in the $a$-direction.}

\ADD{The bulk calculations are performed on a 12$\times$12$\times$12 k-grid in the primitive unit cell and density convergence criterion $1\times10^{-6}$.
For the monolayer setting we performed a slab calculations in the $a-c$ plane and considered
void space in the $b$ direction on a 10$\times$1$\times$10 k-grid. 
In the 2-chain arrangement we employ void space in the $a$ and $b$ direction and 1$\times$1$\times$10 k-grid. The monolayer and 2-chain structure are both converged up to density accuracy of $1\times10^{-4}$. 
}

\section*{Acknowledgement} 
We thank T. Aruga, S. Fujimoto, Y. Fuseya, K. Hirakawa, S. Kasahara, S. Komiyama, T. Osada, \ADD{Y.-J. Song,} K. Sugawara, Y. Sugimoto, K. Takasan, and M.G. Yamada for fruitful discussions.  This work was supported by \ADD{PRESTO (No. JPMJPR225; T.A.) and} CREST (No. JPMJCR19T5;  Y.M. and T.S.) from Japan Science and Technology (JST), Grants-in-Aid for Scientific Research (KAKENHI)  (Nos. 18H05227, 18H03680, 18H01180,21K13881)  and Grant-in-Aid for Scientific Research on innovative areas ‘Quantum Liquid Crystals’ (No. JP19H05824) from Japan Society for the Promotion of Science (JSPS) (T.S.). \ADD{A.R. and R.V. gratefully acknowledge support by the Deutsche Forschungsgemeinschaft (DFG, German Research
Foundation) for funding through TRR 288—422213477 (project A05).}

\section*{Data Availability Statement}
All data needed to evaluate the conclusions in the paper are present in the paper and/or the Supplementary Materials.

\section*{Author contributions} 
T. A. and Y.M. conceived and supervised the study.  T.A., L.P. and S.A. performed STM measurements.  T.A., I.T., T.O., H.M., S.S., Y.Ka. and T.T. synthesized thin films.  T.A., L.P., Y.Ko. and  Y.M.  analyzed the data with theoretical inputs from M.I. and S.-i.S. \ADD{A.R. and R.V. performed the density functional theory calculations.}  T.A., T.S. and Y.M. prepared the manuscript with inputs from Y.Ko., M.I. and S.-i.S. 

\section*{Competing interests}
All authors declare that they have no competing interests. 


\begin{thebibliography}{10}
	
	\bibitem{liu2013top}
	X.~Liu, T.~Xu, X.~Wu, Z.~Zhang, J.~Yu, H.~Qiu, J.-H. Hong, C.-H. Jin, J.-X. Li,
	X.-R. Wang, {\it et~al.\/}, {Top--down fabrication of sub-nanometre
		semiconducting nanoribbons derived from molybdenum disulfide sheets}.
	\newblock {\it Nat. Commun.\/} {\bf 4}, 1--6 (2013).
	
	\bibitem{lin2014flexible}
	J.~Lin, O.~Cretu, W.~Zhou, K.~Suenaga, D.~Prasai, K.~I. Bolotin, N.~T. Cuong,
	M.~Otani, S.~Okada, A.~R. Lupini, {\it et~al.\/}, Flexible metallic nanowires
	with self-adaptive contacts to semiconducting transition-metal dichalcogenide
	monolayers.
	\newblock {\it Nat. Nanotechnol.\/} {\bf 9}, 436--442 (2014).
	
	\bibitem{dyck2019atom}
	O.~Dyck, M.~Ziatdinov, D.~B. Lingerfelt, R.~R. Unocic, B.~M. Hudak, A.~R.
	Lupini, S.~Jesse, S.~V. Kalinin, Atom-by-atom fabrication with electron
	beams.
	\newblock {\it Nat. Rev. Mater.\/} {\bf 4}, 497--507 (2019).
	
	\bibitem{holmes2000control}
	J.~D. Holmes, K.~P. Johnston, R.~C. Doty, B.~A. Korgel, Control of thickness
	and orientation of solution-grown silicon nanowires.
	\newblock {\it Science\/} {\bf 287}, 1471--1473 (2000).
	
	\bibitem{mallet2008growth}
	J.~Mallet, M.~Molinari, F.~Martineau, F.~Delavoie, P.~Fricoteaux, M.~Troyon,
	Growth of silicon nanowires of controlled diameters by electrodeposition in
	ionic liquid at room temperature.
	\newblock {\it Nano Lett.\/} {\bf 8}, 3468--3474 (2008).
	
	\bibitem{kibsgaard2008atomic}
	J.~Kibsgaard, A.~Tuxen, M.~Levisen, E.~L{\ae}gsgaard, S.~Gemming, G.~Seifert,
	J.~V. Lauritsen, F.~Besenbacher, {Atomic-scale structure of Mo$_6$S$_6$
		nanowires}.
	\newblock {\it Nano Lett.\/} {\bf 8}, 3928--3931 (2008).
	
	\bibitem{wagner1964vapor}
	A.~R. Wagner, S.~W. Ellis, Vapor-liquid-solid mechanism of single crystal
	growth.
	\newblock {\it Appl. Phys. Lett.\/} {\bf 4}, 89--90 (1964).
	
	\bibitem{johansson2011recent}
	J.~Johansson, K.~A. Dick, Recent advances in semiconductor nanowire
	heterostructures.
	\newblock {\it CrystEngComm\/} {\bf 13}, 7175--7184 (2011).
	
	\bibitem{garnett2019introduction}
	E.~Garnett, L.~Mai, P.~Yang, Introduction: 1d nanomaterials/nanowires.
	\newblock {\it Chem. Rev.\/} {\bf 119}, 8955--8957 (2019).
	
	\bibitem{whang2003large}
	D.~Whang, S.~Jin, Y.~Wu, C.~M. Lieber, Large-scale hierarchical organization of
	nanowire arrays for integrated nanosystems.
	\newblock {\it Nano Lett.\/} {\bf 3}, 1255--1259 (2003).
	
	\bibitem{liu2010mesostructured}
	J.-W. Liu, J.-H. Zhu, C.-L. Zhang, H.-W. Liang, S.-H. Yu, Mesostructured
	assemblies of ultrathin superlong tellurium nanowires and their
	photoconductivity.
	\newblock {\it J. Am. Chem. Soc.\/} {\bf 132}, 8945--8952 (2010).
	
	\bibitem{kitaev2006anyons}
	A.~Kitaev, Anyons in an exactly solved model and beyond.
	\newblock {\it Ann. Phys.\/} {\bf 321}, 2--111 (2006).
	
	\bibitem{giamarchi2003quantum}
	T.~Giamarchi, {\it {Quantum physics in one dimension}\/}, vol. 121 (Clarendon
	press, 2003).
	
	\bibitem{tsvelik2013majorana}
	A.~Tsvelik, {Majorana fermion realization of a two-channel Kondo effect in a
		junction of three quantum Ising chains}.
	\newblock {\it Phys. Rev. Lett.\/} {\bf 110}, 147202 (2013).
	
	\bibitem{TuringOriginal}
	A.~Turing, The chemical basis of morphogenesis.
	\newblock {\it Phil. Trans. R. Soc. Lond.\/} {\bf B237}, 37--72 (1952).
	
	\bibitem{horvath2009experimental}
	J.~Horv{\'a}th, I.~Szalai, P.~De~Kepper, An experimental design method leading
	to chemical turing patterns.
	\newblock {\it Science\/} {\bf 324}, 772--775 (2009).
	
	\bibitem{kondo1995reaction}
	S.~Kondo, R.~Asai, {A reaction--diffusion wave on the skin of the marine
		angelfish Pomacanthus}.
	\newblock {\it Nature\/} {\bf 376}, 765--768 (1995).
	
	\bibitem{sick2006wnt}
	S.~Sick, S.~Reinker, J.~Timmer, T.~Schlake, {WNT and DKK determine hair
		follicle spacing through a reaction-diffusion mechanism}.
	\newblock {\it Science\/} {\bf 314}, 1447--1450 (2006).
	
	\bibitem{sheth2012hox}
	R.~Sheth, L.~Marcon, M.~F. Bastida, M.~Junco, L.~Quintana, R.~Dahn, M.~Kmita,
	J.~Sharpe, M.~A. Ros, {Hox genes regulate digit patterning by controlling the
		wavelength of a Turing-type mechanism}.
	\newblock {\it Science\/} {\bf 338}, 1476--1480 (2012).
	
	\bibitem{frank1949one}
	F.~C. Frank, J.~H. van~der Merwe, {One-dimensional dislocations. I. Static
		theory}.
	\newblock {\it Proceedings of the Royal Society of London. Series A.
		Mathematical and Physical Sciences\/} {\bf 198}, 205--216 (1949).
	
	\bibitem{volmer1926keimbildung}
	M.~Volmer, A.~Weber, {Keimbildung in {\"u}bers{\"a}ttigten Gebilden}.
	\newblock {\it Zeitschrift f{\"u}r physikalische Chemie\/} {\bf 119}, 277--301
	(1926).
	
	\bibitem{stranski1937theorie}
	I.~N. Stranski, L.~Krastanow, {Zur Theorie der orientierten Ausscheidung von
		Ionenkristallen aufeinander}.
	\newblock {\it Monatshefte f{\"u}r Chemie und verwandte Teile anderer
		Wissenschaften\/} {\bf 71}, 351--364 (1937).
	
	\bibitem{seul1995domain}
	M.~Seul, D.~Andelman, Domain shapes and patterns: the phenomenology of
	modulated phases.
	\newblock {\it Science\/} {\bf 267}, 476--483 (1995).
	
	\bibitem{malescio2003stripe}
	G.~Malescio, G.~Pellicane, Stripe phases from isotropic repulsive interactions.
	\newblock {\it Nat. Mater.\/} {\bf 2}, 97--100 (2003).
	
	\bibitem{de2006columnar}
	A.~De~Candia, E.~Del~Gado, A.~Fierro, N.~Sator, M.~Tarzia, A.~Coniglio,
	Columnar and lamellar phases in attractive colloidal systems.
	\newblock {\it Phys. Rev. E\/} {\bf 74}, 010403 (2006).
	
	\bibitem{cross1993pattern}
	M.~C. Cross, P.~C. Hohenberg, Pattern formation outside of equilibrium.
	\newblock {\it Rev. Mod. Phys.\/} {\bf 65}, 851 (1993).
	
	\bibitem{nicolis1977self}
	G.~Nicolis, Self-organization in nonequilibrium systems.
	\newblock {\it Dissipative Structures to Order through Fluctuations\/} pp.
	339--426 (1977).
	
	\bibitem{kuramoto1984chemical}
	Y.~Kuramoto, Chemical turbulence.
	\newblock {\it Chemical oscillations, waves, and turbulence\/} (Springer,
	1984), pp. 111--140.
	
	\bibitem{castets1990experimental}
	V.~Castets, E.~Dulos, J.~Boissonade, P.~De~Kepper, Experimental evidence of a
	sustained standing turing-type nonequilibrium chemical pattern.
	\newblock {\it Phys. Rev. Lett.\/} {\bf 64}, 2953 (1990).
	
	\bibitem{ouyang1991transition}
	Q.~Ouyang, H.~L. Swinney, Transition from a uniform state to hexagonal and
	striped turing patterns.
	\newblock {\it Nature\/} {\bf 352}, 610--612 (1991).
	
	\bibitem{tan2018polyamide}
	Z.~Tan, S.~Chen, X.~Peng, L.~Zhang, C.~Gao, Polyamide membranes with nanoscale
	turing structures for water purification.
	\newblock {\it Science\/} {\bf 360}, 518--521 (2018).
	
	\bibitem{fuseya2021nanoscale}
	Y.~Fuseya, H.~Katsuno, K.~Behnia, A.~Kapitulnik, Nanoscale turing patterns in a
	bismuth monolayer.
	\newblock {\it Nat. Phys.\/} {\bf 17}, 1031--1036 (2021).
	
	\bibitem{koch1994biological}
	A.~Koch, H.~Meinhardt, Biological pattern formation: from basic mechanisms to
	complex structures.
	\newblock {\it Rev. Mod. Phys.\/} {\bf 66}, 1481 (1994).
	
	\bibitem{lacroix2011introduction}
	C.~Lacroix, P.~Mendels, F.~Mila, {\it Introduction to frustrated magnetism:
		materials, experiments, theory\/}, vol. 164 (Springer Science \& Business
	Media, 2011).
	
	\bibitem{kasahara2018majorana}
	Y.~Kasahara, T.~Ohnishi, Y.~Mizukami, O.~Tanaka, S.~Ma, K.~Sugii, N.~Kurita,
	H.~Tanaka, J.~Nasu, Y.~Motome, {\it et~al.\/}, {Majorana quantization and
		half-integer thermal quantum Hall effect in a Kitaev spin liquid}.
	\newblock {\it Nature\/} {\bf 559}, 227--231 (2018).
	
	\bibitem{yokoi2021half}
	T.~Yokoi, S.~Ma, Y.~Kasahara, S.~Kasahara, T.~Shibauchi, N.~Kurita, H.~Tanaka,
	J.~Nasu, Y.~Motome, C.~Hickey, {\it et~al.\/}, {Half-integer quantized
		anomalous thermal Hall effect in the Kitaev material candidate
		$\alpha$-RuCl$_3$}.
	\newblock {\it Science\/} {\bf 373}, 568--572 (2021).
	
	\bibitem{biswas2019electronic}
	S.~Biswas, Y.~Li, S.~M. Winter, J.~Knolle, R.~Valent{\'\i}, {Electronic
		Properties of $\alpha$-RuCl$_3$ in Proximity to Graphene}.
	\newblock {\it Physical Review Letters\/} {\bf 123}, 237201 (2019).
	
	\bibitem{fu2022transition}
	L.~Fu, C.~Shang, S.~Zhou, Y.~Guo, J.~Zhao, {Transition metal halide nanowires:
		A family of one-dimensional multifunctional building blocks}.
	\newblock {\it Applied Physics Letters\/} {\bf 120}, 023103 (2022).
	
	\bibitem{ali2014large}
	M.~N. Ali, J.~Xiong, S.~Flynn, J.~Tao, Q.~D. Gibson, L.~M. Schoop, T.~Liang,
	N.~Haldolaarachchige, M.~Hirschberger, N.~P. Ong, {\it et~al.\/}, {Large,
		non-saturating magnetoresistance in WTe$_2$}.
	\newblock {\it Nature\/} {\bf 514}, 205--208 (2014).
	
	\bibitem{shang2020atomic}
	C.~Shang, L.~Fu, S.~Zhou, J.~Zhao, {Atomic Wires of transition metal
		chalcogenides: A family of 1D materials for flexible electronics and
		spintronics}.
	\newblock {\it JACS Au\/} {\bf 1}, 147--155 (2020).
	
	\bibitem{Fletcher_1967}
	J.~M. Fletcher, W.~E. Gardner, A.~C. Fox, G.~Topping, {X-Ray, infrared, and
		magnetic studies of $\alpha$- and $\beta$-ruthenium trichloride}.
	\newblock {\it J. Chem. Soc. A\/} pp. 1038--1045 (1967).
	
	\bibitem{Kresse_VASP}
	G.~Kresse, J.~Hafner, Ab initio molecular dynamics for liquid metals.
	\newblock {\it Phys. Rev. B\/} {\bf 47}, 558--561 (1993).
	
	\bibitem{GGA}
	J.~P. Perdew, K.~Burke, M.~Ernzerhof, Generalized gradient approximation made
	simple.
	\newblock {\it Phys. Rev. Lett.\/} {\bf 78}, 1396--1396 (1997).
	
	\bibitem{Dudarev_LDA+U}
	S.~L. Dudarev, G.~A. Botton, S.~Y. Savrasov, C.~J. Humphreys, A.~P. Sutton,
	{Electron-energy-loss spectra and the structural stability of nickel oxide:
		An LSDA+U study}.
	\newblock {\it Phys. Rev. B\/} {\bf 57}, 1505--1509 (1998).
	
	\bibitem{grimme_vdw}
	S.~Grimme, {Semiempirical GGA-type density functional constructed with a
		long-range dispersion correction}.
	\newblock {\it Journal of Computational Chemistry\/} {\bf 27}, 1787-1799
	(2006).
	
	\bibitem{fplo}
	K.~Koepernik, H.~Eschrig, Full-potential nonorthogonal local-orbital
	minimum-basis band-structure scheme.
	\newblock {\it Phys. Rev. B\/} {\bf 59}, 1743--1757 (1999).
	
	\bibitem{Koepernik_2009_atomic_limit}
	E.~R. Ylvisaker, W.~E. Pickett, K.~Koepernik, {Anisotropy and magnetism in the
		$\text{LSDA}+\text{U}$ method}.
	\newblock {\it Phys. Rev. B\/} {\bf 79}, 035103 (2009).
	
	\bibitem{kaib2022electronic}
	D.~A.~S. Kaib, K.~Riedl, A.~Razpopov, Y.~Li, S.~Backes, I.~Mazin, R.~Valent\'i,
	{Electronic and magnetic properties of the RuX$_3$ (X=Cl, Br, I) family: Two
		siblings -- and a cousin?}
	\newblock {\it npj Quantum Mater.\/} {\bf 7} (2022).
	
\end{thebibliography}

\newpage

\renewcommand{\thefigure}{S\arabic{figure}}
\setcounter{figure}{0}

\begin{figure*}[h]
	\centering
	\includegraphics[width=0.7\linewidth]{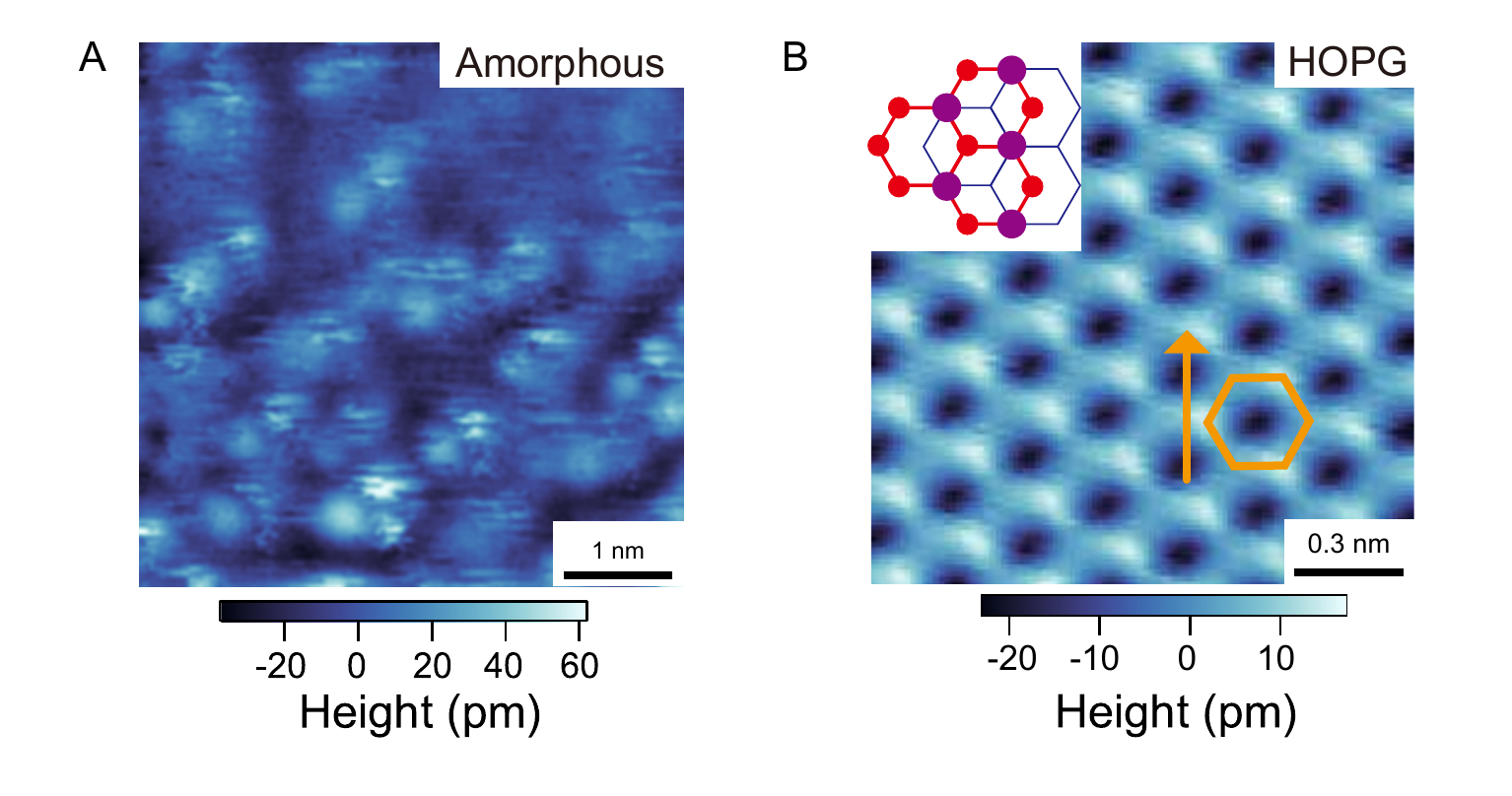}
	\caption{\label{Fig:stripe}
        \textbf{Topographic images of  amorphous and HOPG.}
        \textbf{A,} Typical atomic-scale image of $a$-Ru-Cl, which fills the areas between the wires (dark blue areas in \ADD{Fig.\,\ref{Fig:basic}A and Fig.\,\ref{Fig:basic}B}  taken at \SI{3}{V} and \SI{20}{pA}.  No periodic structure is seen.
        \textbf{B,} Atomic-scale image of HOPG surface.  The orange hexagon and arrow indicate the  carbon honeycomb lattice and the \ADD{chain direction of $\beta$-RuCl$_3$.} . The setpoint conditions are \SI{1}{V} and \SI{100}{pA}.
        } 
\end{figure*}

\begin{figure*}[h]
    \centering
    \includegraphics[width=0.5\linewidth]{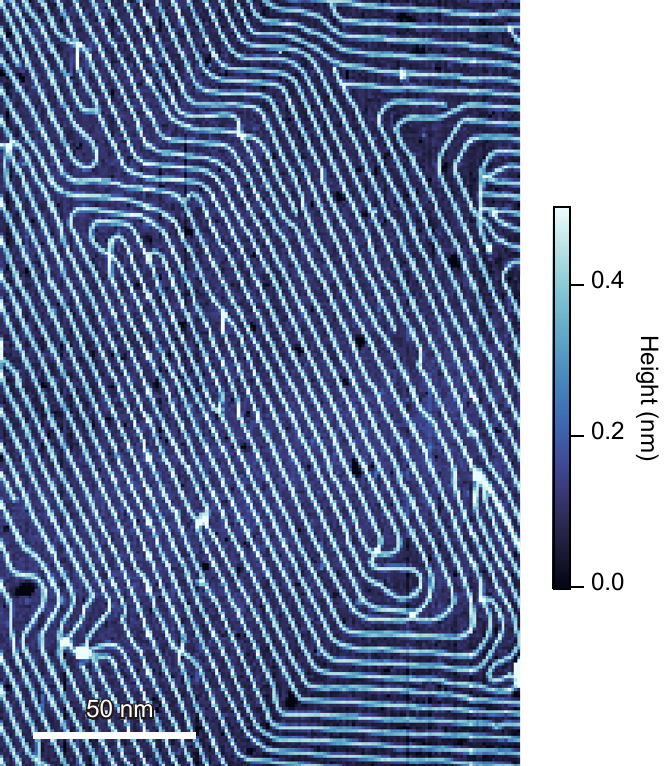}
    \caption{\label{Fig:rawspiral} \textbf{The raw topographic image of the spiral pattern shown in Fig.\,\ref{Fig:evidence}D.} } 
\end{figure*}

\begin{figure*}[h]
    \centering
    \includegraphics[width=0.5\linewidth]{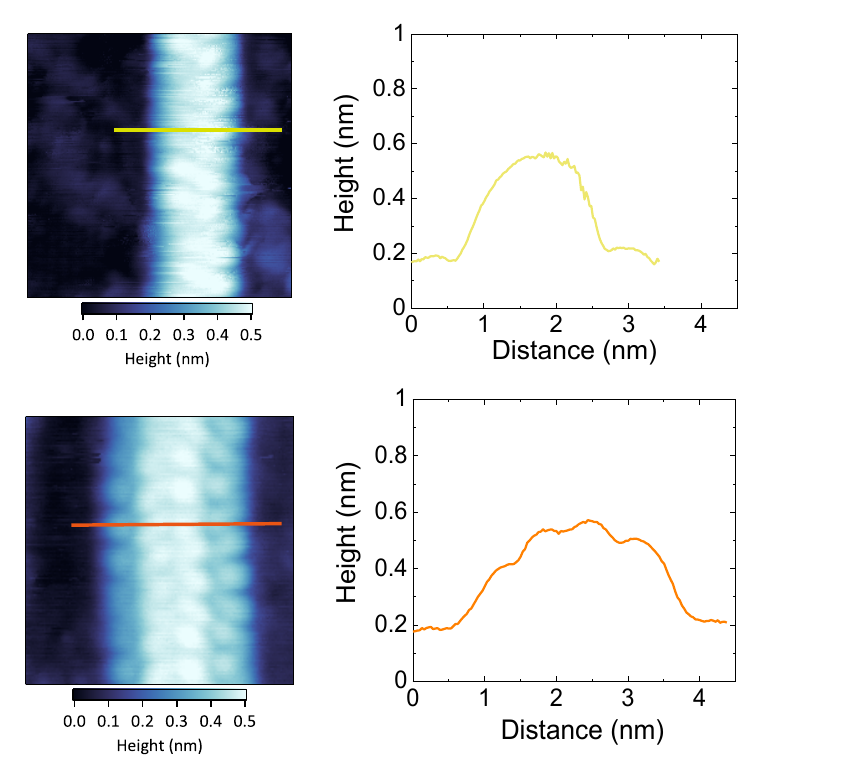}
    \caption{\label{Fig:rawwire}
        \textbf{Atomic wires with different widths.}
        (left) Topographic images of atomic wires with two- (top) and four- (bottom) unit-cell widths, which are grown at \SI{380}{\degreeCelsius} and \SI{400}{\degreeCelsius}, respectively.
        (right) The height profiles along the lines shown in the left panels. The base line is set to the HOPG surface.
        } 
\end{figure*}

\begin{figure*}[h]
    \centering
    \includegraphics[width=0.5\linewidth]{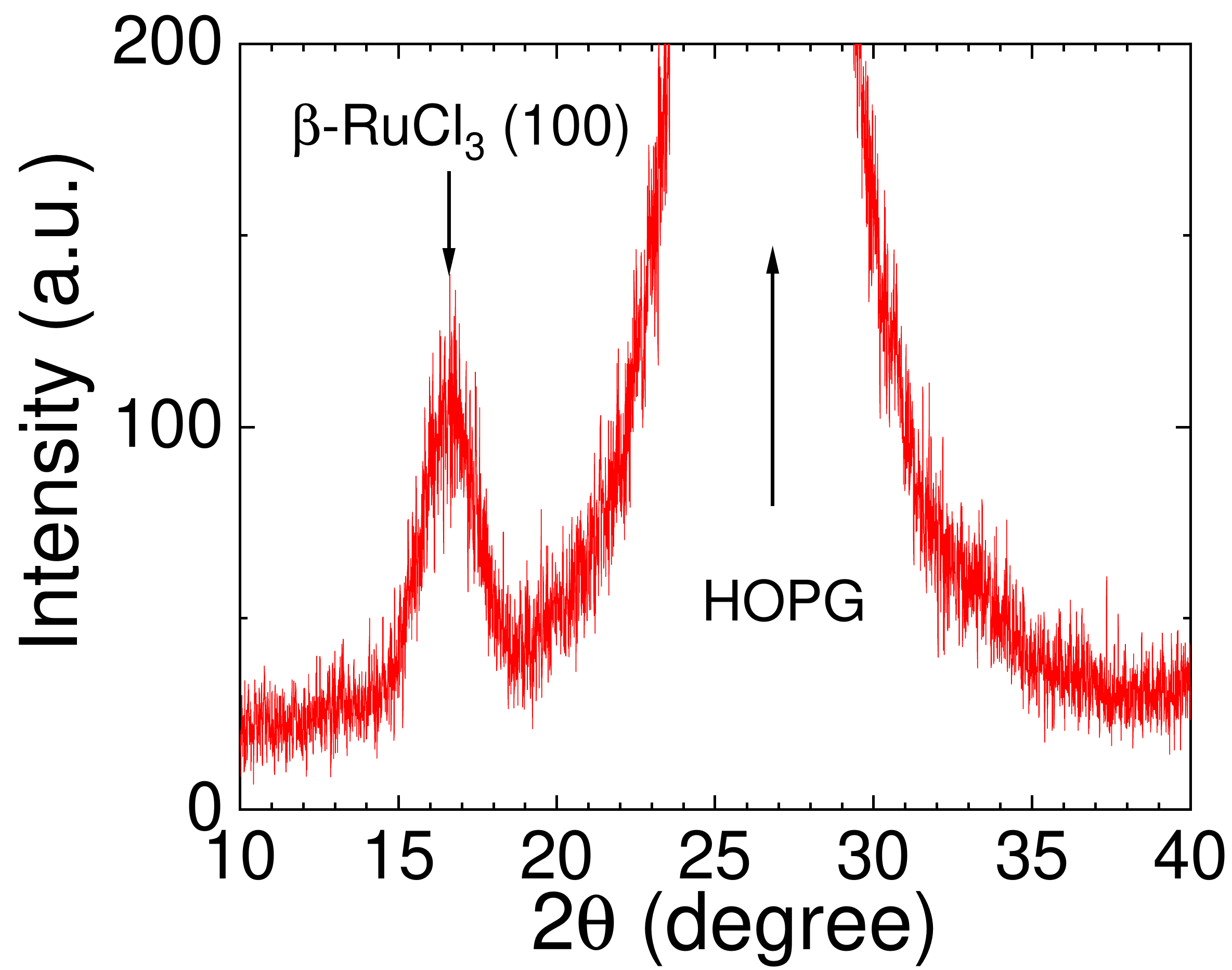}
    \caption{\label{Fig:Xray}
        \textbf{X-ray diffraction pattern of a $\beta$-RuCl$_3$ thin film.}
        A peak at 2$\theta$=\ang{16.7} corresponds to the (100) plane of $\beta$-RuCl$_3$. The thickness of the film is less than \SI{100}{nm}.
        } 
\end{figure*}

\begin{figure*}[h]
    \centering
    \includegraphics[width=0.5\linewidth]{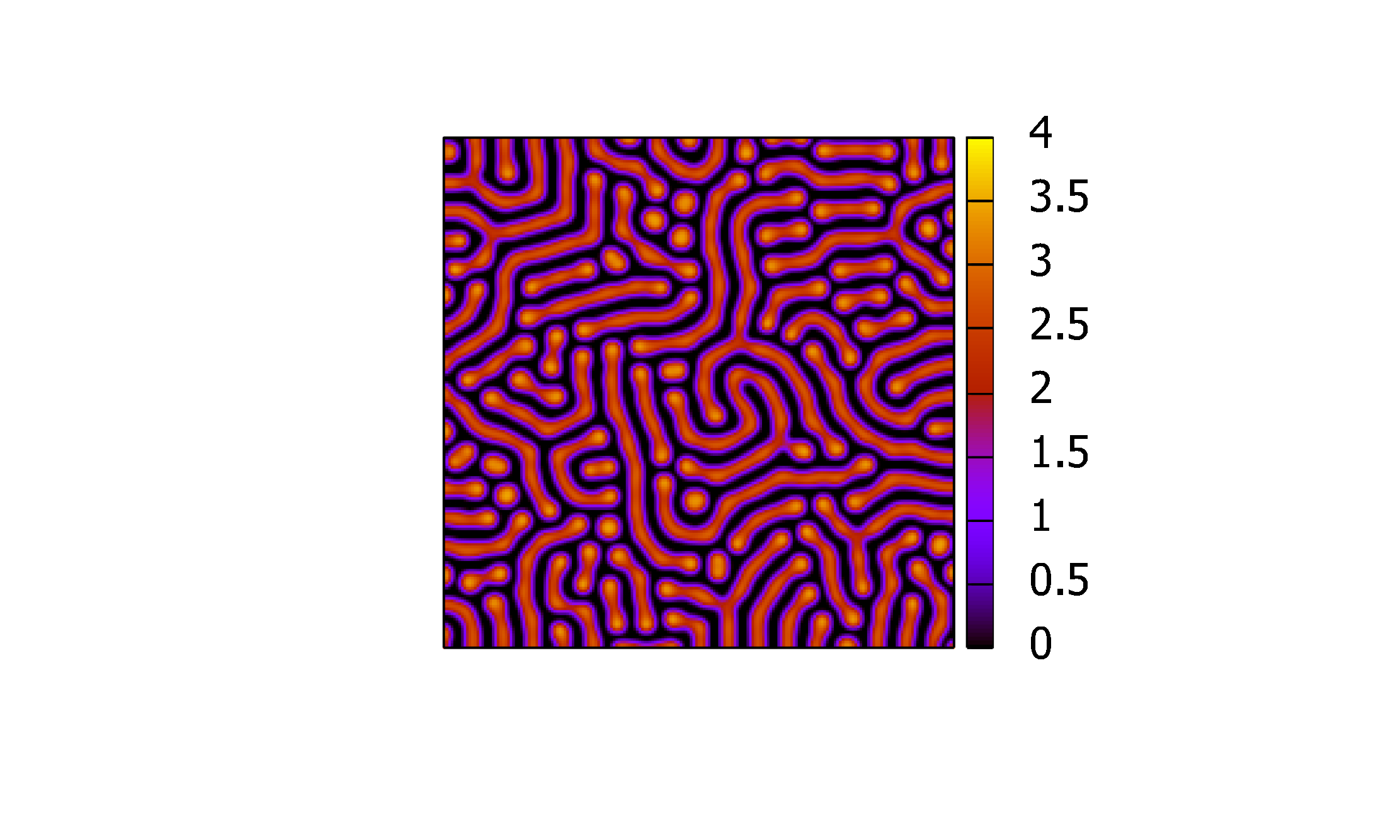}
    \caption{\label{Fig:Simulation}
        \textbf{Simulated Turing pattern.}
        See SI for details. The simulated spiral pattern is also shown in Supplementary Movie\,1.
        } 
\end{figure*}


\begin{figure*}[t]
	\centering
	\includegraphics[width=0.5\linewidth]{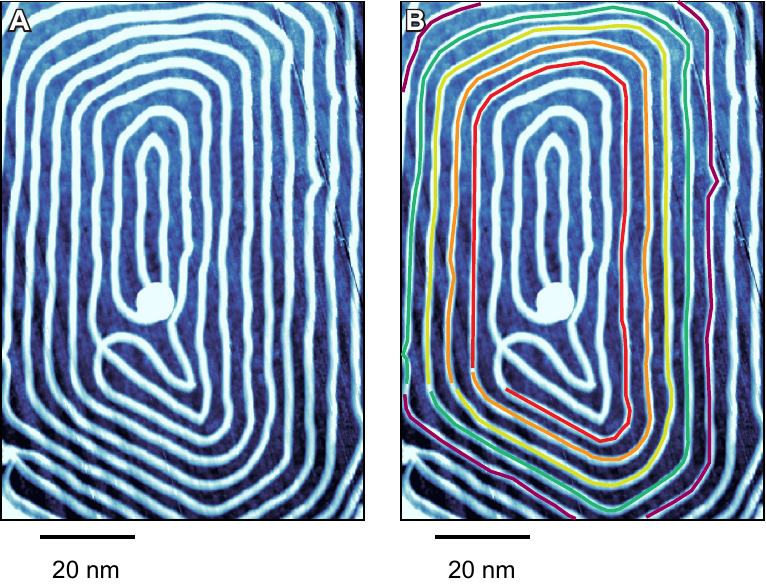}
	\caption{\label{Fig:NewSpiral} 
        \textbf{A spiral pattern with multiple windings.}
        \textbf{A,} The same image as shown in Fig.\,\ref{Fig:evidence}C.
        \textbf{B,} The same image as \textbf{A} with a different color for each lap of the spiral to make the spiral pattern easier to see.
        } 
\end{figure*}


\end{document}


	
\title{Growth of self-integrated atomic quantum wires and junctions of a Mott semiconductor}
\author{ Tomoya\,Asaba$^{1*\dagger}$, Lang\,Peng$^{1*}$, Takahiro\,Ono$^1$, Satoru\,Akutagawa$^{1}$,  Ibuki\,Tanaka$^1$,  Hinako\,Murayama$^{1,2}$, Shota\,Suetsugu$^{1}$, \ADD{Aleksandar\,Razpopov$^{3}$},  Yuichi\,Kasahara$^{1}$, Takahito\,Terashima$^{1}$, Yuhki\,Kohsaka$^1$, Takasada\,Shibauchi$^{4}$, Masatoshi\,Ichikawa$^1$, \ADD{Roser\,Valent\'i$^{3}$,} Shin-ichi\,Sasa$^1$, and  Yuji\,Matsuda$^{1\dagger}$}

\affiliation{
	$^1$Department of Physics, Kyoto University, Kyoto 606-8502, Japan.\\
	$^2$RIKEN Center for Emergent Matter Science, Wako, Saitama 351-0198, Japan\\
	\ADD{$^3$Institut f\"ur Theoretische Physik, Goethe-Universit\"at, 60438 Frankfurt am Main, Germany}\\
	$^4$Department of Advanced Materials Science, University of Tokyo, Kashiwa, Chiba 277-8561, Japan\\
	$^{\dagger}$Correspondence to: {\rm asaba.tomoya.4t@kyoto-u.ac.jp} and {\rm matsuda@scphys.kyoto-u.ac.jp}\\
}
	\date{\today}
	

	\maketitle

\section*{Reaction-diffusion model}

To explain the stripe pattern of $\beta$-RuCl$_3$, we consider a
reaction-diffusion model describing the activator-substrate system
illustrated in 
Fig.\,4a. Let $a$ and $s$ be the concentration fields of activator
(crystalline $\beta$-RuCl$_3$) and substrate (amorphous a-Ru-Cl),
respectively. The important reaction is that a unit constituting the
amorphous becomes a part of the crystal at the rate proportional to
the concentration of the crystal. This auto-catalytic reaction is suppressed
by the depletion of the substrate when its diffusion constant $D_s$ is
much larger than activator's diffusion $D_a$. The mechanism leads to the
Turing instability of a spatially homogeneous state.  As one example, 
the equations for $a$ and $s$ are written as
\begin{align}
  \partial_t  a&= sa-a- 1+D_a (\partial_x^2+\partial_y^2) a,  \label{evol:a}\\
  \partial_t  s&= -sa+2+D_s (\partial_x^2+\partial_y^2) s, \label{evol:s}
\end{align}
where we set $D_a= 1$ and  $D_s=20$ to satisfy the condition $D_s \gg D_a$.
The last condition seems reasonable because the particle hopping rate in the crystalline $\beta$-RuCl$_3$ is much
smaller than that in amorphous a-Ru-Cl. Here, the
isotropic diffusion is assumed, for simplicity. It should be noted
that the form of Eqs. (\ref{evol:a}) and (\ref{evol:s}) was
introduced by Turing. 

By a standard method, it is easily confirmed that the spatially homogeneous
state $(a,s)=(1,2)$ is unstable against periodic perturbations. We then
numerically solve Eqs. (\ref{evol:a}) and (\ref{evol:s}) with the boundary
conditions  $\bv{n}\bv{\nabla}a=\bv{n}\bv{\nabla}s=0$, where $\bv{n}$ is
the unit vector perpendicular to  boundaries. We choose an initial condition
randomly near the homogeneous state. Furthermore, in solving the equations,
we impose the following additional rule: if the concentration $a$ is zero and the increasing rate of $a$ is negative, $a$ remains zero. This special rule was
also mentioned in the paper by Turing. 

An example of the resulting stationary pattern is shown in Extended
Data Fig\, 6, which shows an irregular stripe pattern. 
\ADD{Also, in Supplementary Movie 2, we show the simulated nucleation and propagation of wires. Since thin film growth usually occurs from singularities such as clusters of impurities or steps, the graphite substrate steps are considered to be the nucleus. Considering this, in this simulation, we use Eqs.\,(S1) and (S2) as well, but assume that the initial concentration of $a$ is dense along the left edge of the figure (indicated by warm color). The movie shows that the ordered wire pattern is spontaneously formed.  } 
	
Although this simulation based on simple assumptions should be scrutinized more closely, it captures an essential feature of the observed patterns. 

\section*{Spiral formation of stripes}

Figure 3d shows a spiral of stripes, which is quite unique even
in reaction-diffusion systems. Let us consider a model equation
that exhibits such a pattern. 
First, when the Turing instability occurs, a plane-wave perturbation
with wavenumber vector $\bv{k}$ grows at the rate $\lambda(\bv{k})$. 
The simplest example of $\lambda(\bv{k})$ is given by
\begin{equation}
\lambda(\bm{k})= r-(|\bv{k}|^2-k_c^2)^2, 
\label{linear}
\end{equation}
where $r=0$ corresponds to the onset of instability and 
$k_c$ is the critical wavenumber at $r=0$. 
The exponential growth due to the instability is
saturated by a non-linear effect, while its detailed mechanism depends
on the system under study. Here, we assume that the activator is
described by a two-component field associated with a crystal
structure. Let $W$ be the complex field corresponding to the
two-component field. The equation for $W$ is then written as
\begin{align}
  \partial_t W&=
  (1-(\partial_x^2+\partial_y^2 + 1)^2)W - (1+ic) |W|^2W,
\label{c-SH}
\end{align}
where we have set $(r,k_c)=(1,1)$ in Eq. (\ref{linear}),
and the last term represents 
non-linear saturation. Equation (\ref{c-SH}) with $c=0$ is called the
Swift-Hohenberg equation, 
which describes periodic pattern formations in many systems 
including the Rayleigh-B\'enard convection. The important property
of the Swift-Hohenberg equation is that $W$ evolves with decreasing
a free energy (or a potential function). It has been known that spirals
are never observed in such potential systems. Now, for the case $c \not =0$, 
the system does not possess a potential, and thus spirals may be formed.
Supplementary Movie 1 displays an example of the time evolution of the real
part of $W$ for $c=1$.

{\bf Supplementary Movie 1}
An example of the time evolution of the real part of $W$ for $c=1$. See section {\bf Spiral formation of stripes} for details.

{\bf Supplementary Movie 2}
An example of the time evolution of  $a$ in Eqs.\,(S1) and (S2). See  section {\bf Reaction-diffusion model} for details.


\renewcommand{\thefigure}{S\arabic{figure}}
\setcounter{figure}{0}

\begin{figure*}[h]
	\centering
	\includegraphics[width=0.7\linewidth]{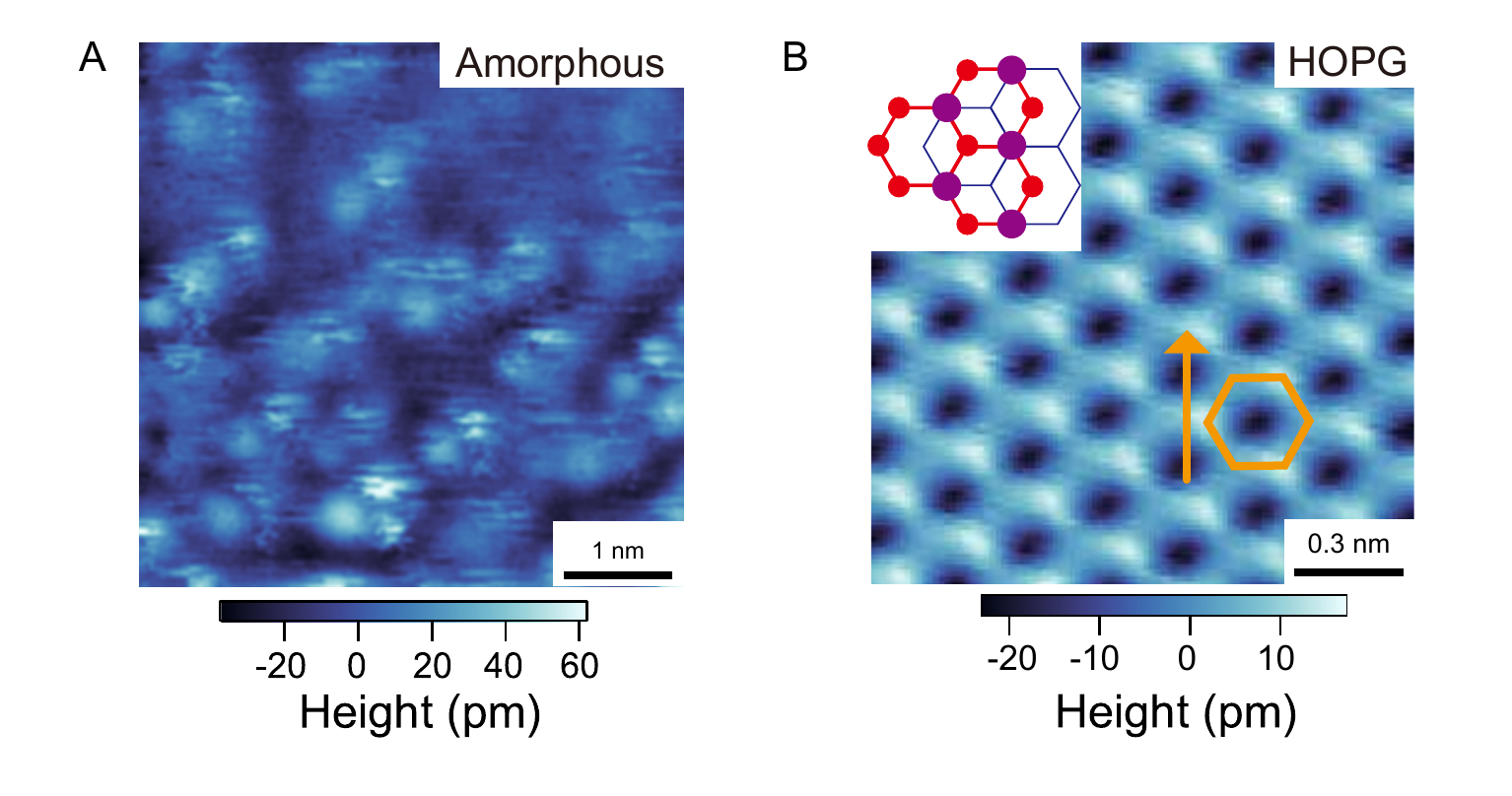}
	\caption{\label{Fig:stripe}
		\textbf{Topographic images of  amorphous and HOPG.}
		\textbf{A,} Typical atomic-scale image of $a$-Ru-Cl, which fills the areas between the wires (dark blue areas in \ADD{Fig.\,1A and Fig.\,1B}  taken at \SI{3}{V} and \SI{20}{pA}.  No periodic structure is seen.
		\textbf{B,} Atomic-scale image of HOPG surface.  The orange hexagon and arrow indicate the  carbon honeycomb lattice and the \ADD{chain direction of $\beta$-RuCl$_3$.} . The setpoint conditions are \SI{1}{V} and \SI{100}{pA}.		
	} 
\end{figure*}

\begin{figure*}[h]
	\centering
	\includegraphics[width=0.5\linewidth]{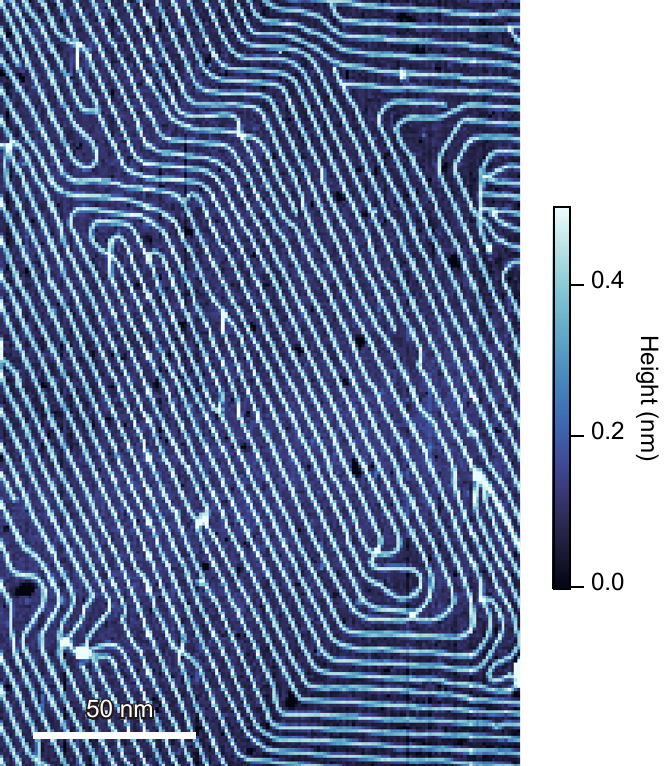}
	\caption{\label{Fig:rawspiral} \textbf{The raw topographic image of the spiral pattern shown in Fig.\,6D.} } 
\end{figure*}

\begin{figure*}[h]
	\centering
	\includegraphics[width=0.5\linewidth]{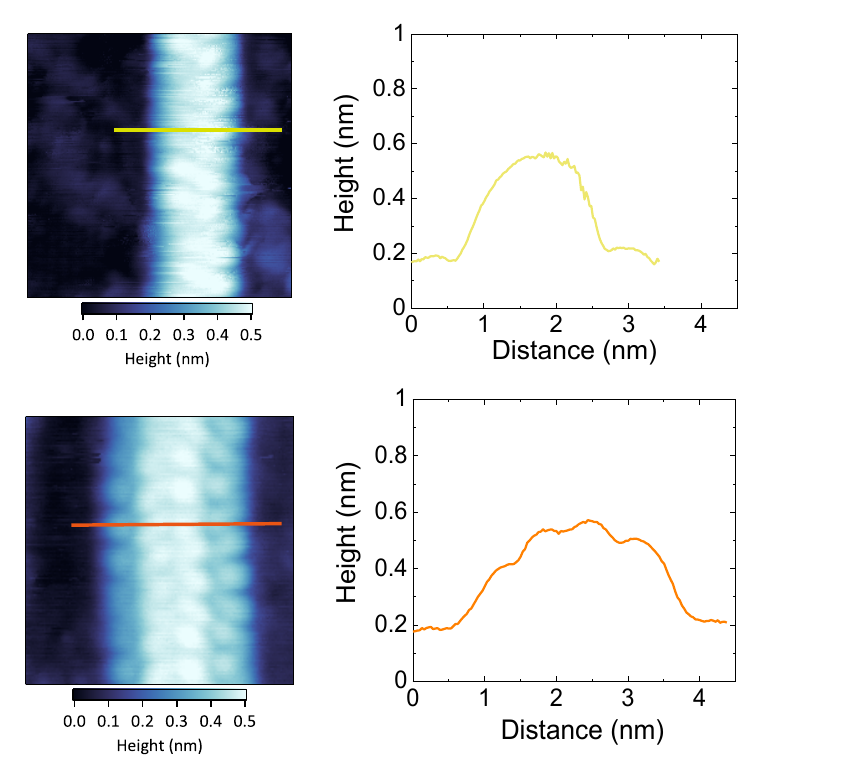}
	\caption{\label{Fig:rawwire}
		\textbf{Atomic wires with different widths.}
		(left) Topographic images of atomic wires with two- (top) and four- (bottom) unit-cell widths, which are grown at \SI{380}{\degreeCelsius} and \SI{400}{\degreeCelsius}, respectively.
		(right) The height profiles along the lines shown in the left panels. The base line is set to the HOPG surface.
	} 
\end{figure*}

\begin{figure*}[h]
	\centering
	\includegraphics[width=0.5\linewidth]{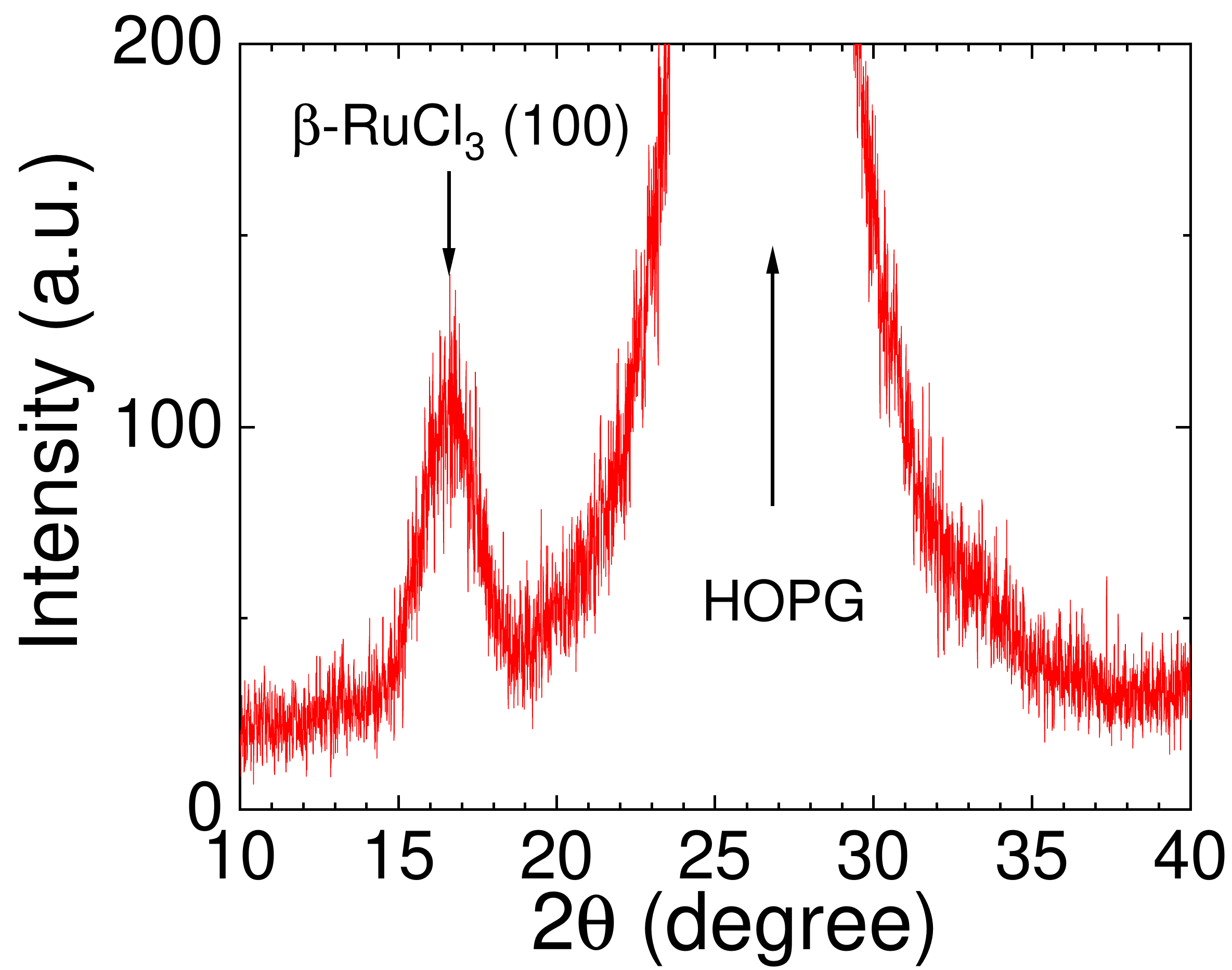}
	\caption{\label{Fig:Xray}
		\textbf{X-ray diffraction pattern of a $\beta$-RuCl$_3$ thin film.}
		A peak at 2$\theta$=\ang{16.7} corresponds to the (100) plane of $\beta$-RuCl$_3$. The thickness of the film is less than \SI{100}{nm}.
	} 
\end{figure*}

\begin{figure*}[h]
	\centering
	\includegraphics[width=0.5\linewidth]{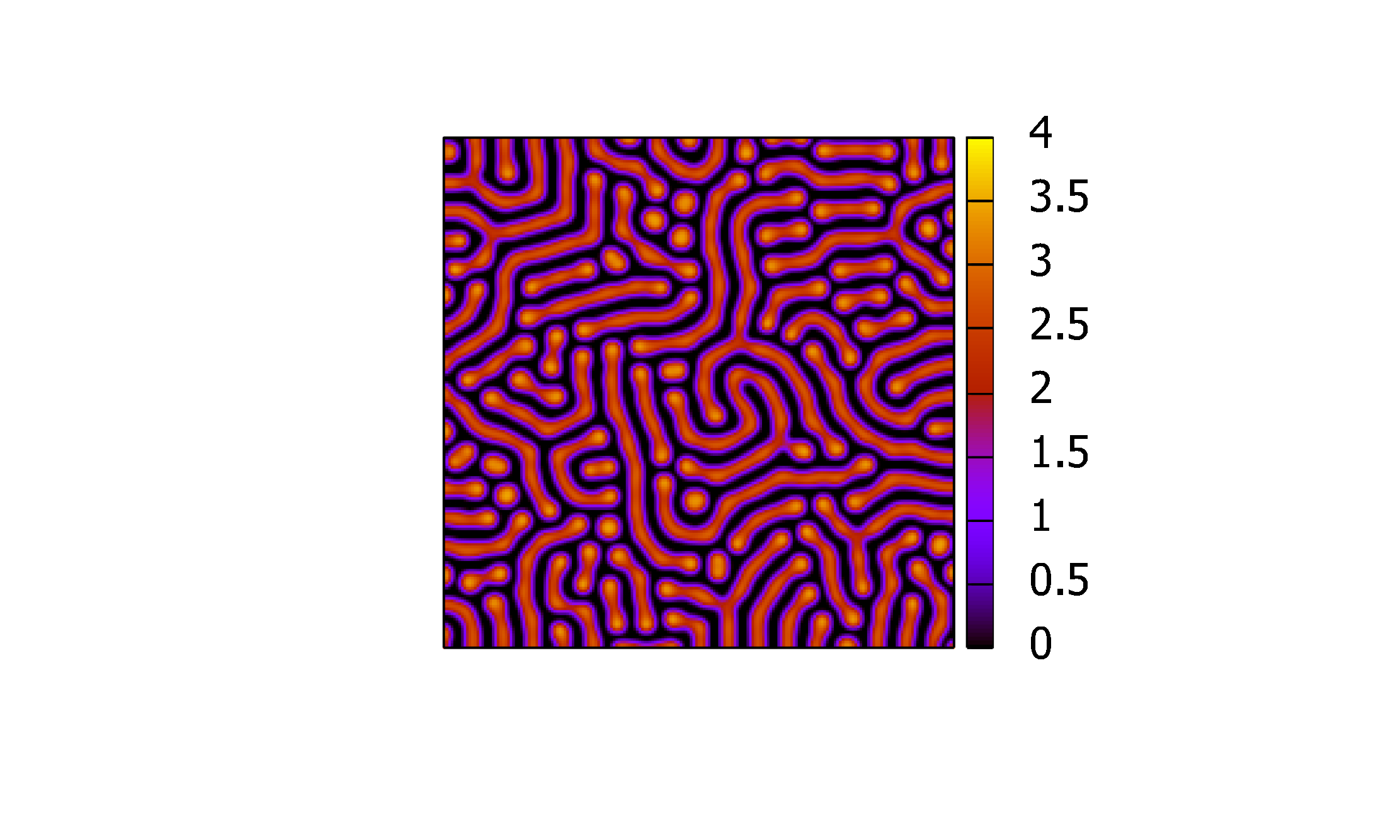}
	\caption{\label{Fig:Simulation}
		\textbf{Simulated Turing pattern.}
		See SI for details. The simulated spiral pattern is also shown in Supplementary Movie\,1.
	} 
\end{figure*}

\begin{figure*}[t]
	\centering
	\includegraphics[width=0.5\linewidth]{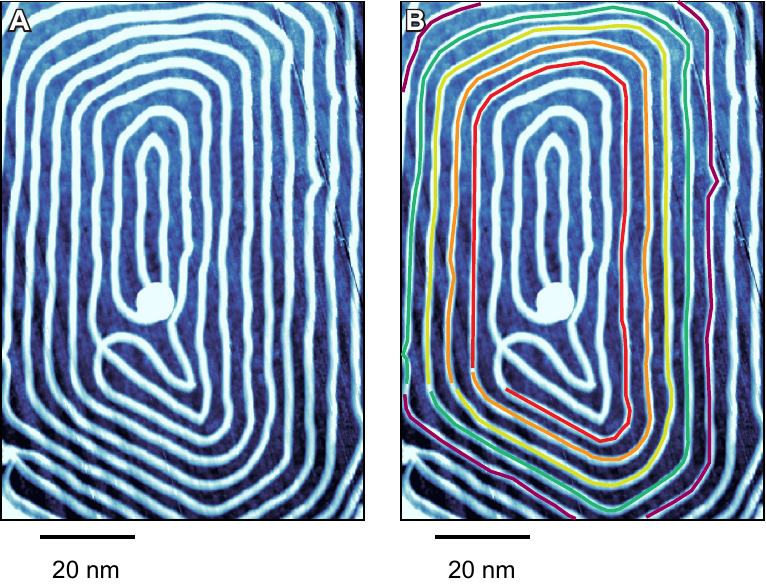}
	\caption{\label{Fig:NewSpiral} 
		\textbf{A spiral pattern with multiple windings.}
		\textbf{A,} The same image as shown in Fig.\,6C.
		\textbf{B,} The same image as \textbf{A} with a different color for each lap of the spiral to make the spiral pattern easier to see.
	} 
\end{figure*}